\documentclass[aps,prx,twocolumn,10pt,groupedaddress]{revtex4-2}
\usepackage[utf8]{inputenc}
\usepackage{etoolbox}
\usepackage{hhline}
\usepackage{array}
\usepackage{amssymb,amsmath,amsthm,amsfonts}
\usepackage{graphicx}
\usepackage{latexsym}
\usepackage{bm}
\usepackage[]{hyperref}
\usepackage{cleveref}
\usepackage{nccmath}	
\usepackage{color}
\usepackage{ulem}
\usepackage{subfigure}
\usepackage{float}
\usepackage{color}
\usepackage{ulem}
\usepackage[utf8]{inputenc}

\allowdisplaybreaks[2]

\newcommand*{\dif}{\mathop{}\!\mathrm{d}}

\begin{document}
	\title{Instability in charged Gauss–Bonnet–de Sitter black holes}
	
	\author{Rong-Gen Cai$^{1,2,3}$}
	\email{cairg@itp.ac.cn}
	\author{Li Li$^{1,2,3}$,}
	\email{liliphy@itp.ac.cn}
	\author{Hao-Tian Sun$^{1,2}$}
	\email{sunhaotian@itp.ac.cn}
	\affiliation{$^1$CAS Key Laboratory of Theoretical Physics, Institute of Theoretical Physics, Chinese Academy of Sciences, Beijing 100190, China}
	\affiliation{$^2$School of Physical Sciences, University of Chinese Academy of Sciences, No.19A Yuquan Road, Beijing 100049, China}
	\affiliation{$^3$School of Fundamental Physics and Mathematical Sciences, Hangzhou Institute for Advanced Study, UCAS, Hangzhou 310024, China}
	\begin{abstract}
		{We study the instability of the charged Gauss-Bonnet de Sitter black holes under gravito-electromagnetic perturbations. We adopt two criteria to search for an instability of the scalar type perturbations, including the local instability criterion based on the $AdS_2$ Breitenl\"{o}hner-Freedman (BF) bound at extremality and the dynamical instability via quasinormal modes by full numerical analysis. We uncover the gravitational instability in five spacetime dimensions and above, and construct the complete parameter space in terms of the ratio of event and cosmological horizons and the Gauss-Bonnet coupling. We show that the BF bound violation is a sufficient but not necessary condition for the presence of dynamical instability. While the physical origin of the instability without the Gauss-Bonnet term has been argued to be from the $AdS_2$ BF bound violation, our analysis suggests that the BF bound violation can not account for all physical origin of the instability for the charged Gauss-Bonnet black holes.}
	\end{abstract}

	\maketitle

	\section{Introduction}
	
	Black holes are one of the most fantastic predictions of Einstein's theory of General Relativity and have been widely investigated both in theory and experiment. While the Einstein equation provides a good description for the evolutions of spacetime and the interaction with matter, it is intrinsically complicated and difficult to solve, making it harder to understand the properties of a black hole. A good starting point is to consider perturbation theory with the background geometry fixed. It is not only an important approach to understanding the linear stabilities or instabilities of systems, but also a useful tool for finding new black hole solutions.
	
	One of the interesting case is the existence of a linear instability for the Reissner-Nordstr\"{o}m de Sitter (RNdS) black hole with a positive cosmological term. While it has been proved that the RNdS background is linear-mode stable for spacetime dimension $d=4, 5$~\cite{Kodama:2003kk}, the authors of~\cite{Konoplya:2008au,Konoplya:2013sba} found that the RNdS black hole can be unstable to gravitational perturbations in seven dimensions and above ($d\geqslant 7$). More recently, the authors of~\cite{Dias:2020ncd} provided strong numerical evidence for the existence of a dynamical instability triggered by the scalar-type gravitational perturbations in $d=6$. The physical origin of such dynamical instability was argued to be due to the violation of the $AdS_2$ BF  bound in the near-horizon limit of extremal RNdS~\cite{Dias:2020ncd}, shedding some light on the nature of the instability. In particular, at extremality, once there exist BF bound violating modes at a finite range of parameter space, the system is unstable in the full range of the parameter space. In some sense, the observation of~\cite{Dias:2020ncd} comes as a surprise: a local instability (the BF bound violation near the extremal horizon regime) is able to account for the dynamical instability which depends on the global geometry of the background as well as the boundary conditions.
	
	In order to understand how general the phenomena found in~\cite{Dias:2020ncd} and to better understand the nature of the instability of charged black holes, it is valuable to classify other extensions of the analysis, in particular, for other gravity theories and spacetime dimensions. A natural generalization is the Einstein Gauss-Bonnet (GB) gravity. Higher derivative curvature terms naturally occur in many occasions, in particular in the low-energy effective action of superstring theory. Of particular interest is the GB term which is the most general quadratic expression in curvature resulting in field equations containing no more than second derivatives of the metric and thus no ghost-like mode. The thermodynamic properties of the static vacuum GB black holes in de Sitter spacetime were studied in~\cite{Cai:2003gr}. The dynamical instability for $d\geqslant 5$ was uncovered from the time-domain analysis for these black holes at large values of the GB coupling and cosmological constant~\cite{Cuyubamba:2016cug}, which was supposed to have the same origin as the instability of higher dimensional RNdS black holes found in~\cite{Konoplya:2008au,Konoplya:2013sba}. More recently, the instability of the Gauss-Bonnet Reissner-Nordstr\"{o}m de Sitter (GB-RNdS) black hole at large $d$ limit was studied in~\cite{Chen:2017hwm}, where the effective equations at large $d$ were introduced to describe the nonlinear dynamical deformations of the black holes. Nevertheless, a comprehensive study on the instability of the GB-RNdS black hole has not been done yet. Compared with the RNdS case, in the GB-RNdS black hole there is an additional knob, \emph{i.e.} the GB coefficient that can be tuned to test if the physical origin of the instability given in~\cite{Dias:2020ncd} is robust or not. In the present work, we will investigate the effect of the GB coupling on the instability of the GB-RNdS black hole. 
	
	For the GB-RNdS black hole, according to the transformation law under rotations on the sphere, the linear perturbations can be classified into tensor, vector and scalar types. Each of them can be treated independently from the others. Following~\cite{Dias:2020ncd}, we shall consider the scalar type perturbations, for which the master equations are given by two coupled second order differential equations. The equations effectively behave as massive scalar fields in an $AdS_2$ background which is the near-horizon limit of the extremal GB-RNdS black hole. We shall adopt two instability criteria. The first one is the Durkee-Reall instability criterion based on the $AdS_2$ BF bound of the near-horizon limit at extremality. The Durkee-Reall criterion~\cite{Durkee:2010ea} conjectures that when the near-horizon effective mass violates the $AdS_2$ BF bound, then the full spacetime geometry is linear unstable. The other one is the quasi-normal mode (QNM) spectrum by solving the master equations between the event and cosmological horizons. The presence of a QNM frequency with a positive imaginary part signifies a dynamical instability and the magnitude of its imaginary part sets the timescale of the instability. Compared with the RNdS case~\cite{Dias:2020ncd}, we shall uncover some novel features for the GB-RNdS black hole. We shall show that the instability of the black hole depends not only on its size and the spacetime dimension, but also significantly on the GB term. For example, we will show that while the standard RNdS black hole which is linear stable in $d=5$ dimensional spacetime~\cite{Kodama:2003kk}, the RNdS black hole with GB coupling in 5 dimensions {can become unstable, depending on the parameters of the black hole}. Our analysis will also suggest that the physical origin of the instability for the GB-RNdS background is not solely due to the violation of the $AdS_2$ BF bound.
	
	The paper is organized as follows. In Section~\ref{sec:setup}, we briefly review the charged Gauss-Bonnet black holes with a positive cosmological constant and give the master equations for the perturbations of the scalar type. In Section~\ref{sec:BF}, we query for the instabilities from the near-horizon $AdS_2$ geometry at extremality. The violation of the $AdS_2$ BF bound provides a criterion for the presence of an instability. A full numerical analysis for the dynamical instabilities is given in Section~\ref{sec:qnm} by searching for the QNMs. {We shall mainly focus on the extremal black holes and give a brief discussion on the non-extremal case.} In Section~\ref{sec:proof}, we prove that the unstable QNMs are purely imaginary.
	We conclude with some discussions in Section~\ref{sec:discussion}. Numerical techniques for QNMs are discussed in Appendix~\ref{app:detail}.

	\section{Background and master equations}\label{sec:setup}
	We consider the Einstein-Gauss-Bonnet gravity with a positive cosmological constant $\Lambda$, coupled with a Maxwell field in $d=n+2$ dimensional spacetime. The action reads
	\begin{align}\label{action}
		S=&\frac{1}{16\pi G}\int\dif^dx\sqrt{-g}\bigg(R-2\Lambda+\bar{\alpha}L_{GB}-\frac{1}{4}F_{\mu\nu}F^{\mu\nu}\bigg),\quad\\
		 \Lambda=&\frac{n(n+1)}{2\bar{L}^2}\,,
	\end{align}
	where the GB term is given by
	\begin{equation}\label{GB}
		L_{GB}=\big(R_{\mu\nu\rho\sigma}R^{\mu\nu\rho\sigma}-4R_{\mu\nu}R^{\mu\nu}+R^2\big)\,,
	\end{equation}
	with $R$, $R_{\mu\nu}$ and $R_{\mu\nu\rho\sigma}$ the Ricci scalar, the Ricci tensor and the Riemann tensor associated with the metric $g_{\mu\nu}$,  respectively. $\bar{L}$ is the de Sitter length scale and $F_{\mu\nu}=\nabla_\mu A_\nu-\nabla_\nu A_\mu$ the strength of the U(1) gauge potential $A_\mu$. The GB term is a topological invariant for $d=4$ and hence does not contribute
	to the gravitational dynamics, thus we consider $d\geqslant 5$. The constant $\bar{\alpha}$ is the GB coefficient and is positive in string theory. Therefore, we shall restrict in this paper to the case $\bar{\alpha}>0$.

	The variation of the action~\eqref{action} with respect to the metric yields the gravity equations
	\begin{equation}\label{EGB eq}
		\mathcal{G}_{\mu\nu}\equiv R_{\mu\nu}-\frac{1}{2}R g_{\mu\nu}+\Lambda g_{\mu\nu}+\bar{\alpha} H_{\mu\nu}=\mathcal{T}_{\mu\nu}\,,
	\end{equation}
	with
	\begin{align}
		H_{\mu\nu}=&2(R R_{\mu\nu}-2R_{\mu\rho}{R^{\rho}}_{\nu}-2R^{\sigma\rho}R_{\mu\sigma\nu\rho}+R_{\mu\alpha\tau\rho}{R_{\nu}}^{\alpha\tau\rho})\nonumber\\
		&-\frac{1}{2}L_{GB} g_{\mu\nu}\,,
	\end{align}
	and
	\begin{equation}
		\mathcal{T}_{\mu\nu}=\frac{1}{2}F_{\mu\rho}{F_\nu}^\rho-\frac{1}{8}F_{\alpha\beta}F^{\alpha\beta} g_{\mu\nu}\,,
	\end{equation}
	the energy-momentum tensor of the Maxwell field. The equation of motion for $A_\mu$ is given by
	\begin{equation}\label{Maxeq}
		\nabla_\mu F^{\mu\nu}=0\,.
	\end{equation}

	The theory allows the GB-RNdS black hole that is given as follows~\cite{Wiltshire:1985us,Cai:2001dz}.
	\begin{equation}\label{RN}
		\begin{split}
			ds^2=&-f(\bar{r})dt^2+\frac{d\bar{r}^2}{f(\bar{r})}+\bar{r}^2 d\Omega_n^2, \quad  A_t=\sqrt{\frac{n}{n-1}}\frac{\bar{Q}}{\bar{r}^{n-1}}\,,\\
			f(\bar{r})=&1+\frac{\bar{r}^2}{2\bar{\alpha}(n-1)(n-2)}\Big(1-\,\\
			&\hspace{1mm}\sqrt{1+\bar{\alpha}(n-1)(n-2)\left[\frac{8\bar{M}}{\bar{r}^{n+1}}-\frac{4\bar{Q}^2}{\bar{r}^{2n}}+\frac{4}{\bar{L}^2}\right]}\Big)\,,
		\end{split}
	\end{equation}
	where $d\Omega_n^2$ is the linear element of a unit radius sphere $S^n$, $\bar{M}$ is the black hole mass and $\bar{Q}$ is related to the charge of the black hole. By taking the limit $\bar{\alpha}\rightarrow 0$, one recovers the Einstein-Maxwell theory as well as the RNdS black hole for which
	\begin{align}\label{BHsolution0}
		f(\bar{r})&=1-\frac{\bar{r}^2}{\bar{L}^2}-\frac{2\bar{M}}{{\bar{r}}^{n-1}}+\frac{\bar{Q}^2}{\bar{r}^{2(n-1)}}, \quad\\ A_t&=\sqrt{\frac{n}{n-1}}\frac{\bar{Q}}{\bar{r}^{n-1}}\,.
	\end{align}
	Therefore, after considering the small $\bar{\alpha}$ limit, our results below should come back to the RNdS case discussed in~\cite{Dias:2020ncd}.
	
	Besides the event horizon $\mathcal{H}^+$ of the GB-RNdS black hole~\eqref{RN}, it has a cosmological horizon $\mathcal{H}_C$ outside $\mathcal{H}^+$ due to the presence of the positive cosmological constant $\Lambda$. For appropriate range parameters, there is also an inner Cauchy horizon $\mathcal{CH}$. We denote the location of the event horizon, cosmological horizon and Cauchy horizon as $\bar{r}_+, \bar{r}_c$ and $\bar{r}_-$, respectively. They are all real positive roots of $f$ with $\bar{r}_-\leqslant \bar{r}_+\leqslant\bar{r}_c$. The temperature of $\mathcal{H}^+$ and $\mathcal{\mathcal{H}_C}$ are respectively given by $\bar{T}_+=\frac{1}{4\pi}\frac{\dif f(\bar{r}_+)}{\dif\bar{r}}$ and $\bar{T}_c=-\frac{1}{4\pi}\frac{\dif f(\bar{r}_c)}{\dif\bar{r}}$.
	
	Note that the system is invariant under the scaling $g\rightarrow \lambda^2g, A\rightarrow \lambda A, \bar{L}\rightarrow \lambda\bar{L}, \bar{\alpha}\rightarrow \lambda^2\bar{\alpha}$ with $\lambda$ a constant. In the following discussion, we shall work with dimensionless quantities in unites of $\bar{r}_c$. More precisely, we introduce
	\begin{align}
		M=\frac{\bar{M}}{r_c^{n-1}},\quad Q=\frac{\bar{Q}}{r_c^{n-1}},\quad r=\frac{\bar{r}}{r_c},\quad\alpha=\frac{\bar{\alpha}}{r_c^2},\quad L=\frac{\bar{L}}{r_c}\,,
	\end{align} 
	such that $\{M,Q,r,\alpha,L\}$ are all dimensionless quantities. For a given $n$, the parameters of the GB-RNdS black hole~\eqref{RN} $\{M,Q,L,r_+, \alpha\}$ are not independent. Denoting $Q_{ext}$ as the charge for extremal black hole for which $\bar{T}_+=0$ and introducing $y_+=r_+/r_c$, one obtains three independent dimensionless parameters $\{y_+,\alpha, q\}$ with $q\equiv Q/Q_{ext}$ the charge ratio. Other parameters can be fixed by $\{y_+,\alpha, q\}$. In particular, one has 
	\begin{align}
		{M}&=\frac{-{Q}^2( y_+^{1-n}-y_+^{n+1})-\alpha  y_+^{n-3}+(\alpha+1)  y_+^{n+1}-y_+^{n-1}}{2 \left(y_+^{n+1}-1\right)}\,,\\
		{L}^2&=\frac{y_+^{n+3} \left(y_+^{n+1}-1\right)}{-{Q}^2 y_+^{n+3}+\alpha  y_+^{2 n}-\alpha  y_+^{n+3}-y_+^{n+3}+y_+^{2 n+2}+{Q}^2 y_+^4}\,,
	\end{align}	
	and 
	\begin{align}
		{Q}^2_{ext}=\frac{{\alpha}  y_+^{2 n-4} \left[4 y_+^{n+1}-n \left(y_+^4-1\right)-y_+^4-3\right]}{y_+^{2 n}+n \left(y_+^{2 n}-2 y_+^{n+1}+1\right)-1}\nonumber\\
		-\frac{y_+^{2 n-2} \left[-2 y_+^{n+1}+n \left(y_+^2-1\right)+y_+^2+1\right]}{y_+^{2 n}+n \left(y_+^{2 n}-2 y_+^{n+1}+1\right)-1}\,.
	\end{align}
	Note that $Q_{ext}$ is independent of ${\alpha}$ when $n=3$. In the following discussion, we shall mainly focus on the extremal black hole for which $Q=Q_{ext}$ ($q=1$) and the independent parameters are $\{y_+,\alpha\}$.
	
	The equations of motion for the linear perturbations $(\delta g_{\mu\nu}, \delta A_\mu)$ around the background~\eqref{RN} can be obtained by taking the variation of the equations~\eqref{EGB eq} and~\eqref{Maxeq}.
	Following the main idea by the authors of~\cite{Kodama:2003kk}, the perturbations of~\eqref{RN} can be classified into different types according to how they transform under diffeomorphisms of $S^n$. Master equations for scalar, vector and tensor modes were obtained for static charged Lovelock black holes in~\cite{Takahashi:2011qda,Takahashi:2012np}\,(the author of~\cite{Takahashi:2011qda,Takahashi:2012np} studied the instability of the static charged Lovelock black holes in asymptotic flat spacetime, \emph{i.e.} $\Lambda=0$ by using the ``S-deformation" approach~\cite{Ishibashi:2003ap,Kodama:2003ck} ). Since we shall generalize the recent discussion of the instability in the RNdS black holes~\cite{Dias:2020ncd} to the GB-RNdS case, we consider the scalar type perturbations, for which the master equations in Fourier space are given by~\cite{Takahashi:2012np}
	\begin{align}\label{Meq}
		\mathcal{H}\begin{bmatrix}
			\phi_{\omega l}\\
			A_{\omega l}
		\end{bmatrix}=\omega^2\begin{bmatrix}
			\phi_{\omega l}\\
			A_{\omega l}
		\end{bmatrix}\,,
	\end{align}
	where
	\begin{align}\label{op_H}
		\mathcal{H}=-\frac{d^2}{dr_*^2}+\begin{bmatrix}
			V_g& V_c\\
			V_c& V_{em}
		\end{bmatrix}\,,
	\end{align}
	and
	\begin{equation}
		\begin{split}
			V_g(r)&=\kappa_s\frac{f}{nr}\left(4(\kappa_s-n)\frac{T^{\prime}}{{\mathcal A}T}-\frac{T^{\prime\prime}}{T^{\prime}}\right)\\
			\ &\hspace{0.5cm}+\frac{2n(r^nE)^2f^2}{{\mathcal A}Tr^n}\left(\ln\left(\frac{fT^{\prime}}{r^{n-2}({\mathcal A}T)^2}\right)\right)^{\prime}\\
			&\hspace{0.5cm}+\frac{{\mathcal A}T}{r\sqrt{T^{\prime}}}f\partial_r\left(f\partial_r\frac{r\sqrt{T^{\prime}}}{{\mathcal A}T}\right)\,,\\
			V_{c}(r)&=\sqrt{\frac{\kappa_s-n}{n}}\frac{\sqrt{T^{\prime}}}{r^{n/2}{\mathcal A}}\Big[-\kappa_s\frac{4(r^nE)f}{rT}\\
			&\hspace{0.5cm}+\frac{2n(r^nE)f^2}{T}\left(\ln\left(\frac{fT^{\prime}}{r^{n-2}({\mathcal A}T)^2}\right)\right)^{\prime}\Big]\,,\\
			V_{em}(r)&=fr^{(n-2)/2}\partial_r(f\partial_r\frac{1}{r^{(n-2)/2}})+\kappa_s\frac{f}{r^2}\left(1+\frac{4(Er^n)^2}{{\mathcal A}Tr^{n-1}}\right)\\
			\ &\hspace{0.3cm}+\frac{2n(r^nE)^2f^2}{r^n{\mathcal A}T}\left(\ln\left(\frac{r^{2n-2}({\mathcal A}T)^2}{f}\right)\right)^{\prime}\,,
			\label{V}
		\end{split}
	\end{equation}
	with
	\begin{equation}\label{myT}
		\begin{split}
			&\kappa_s=l(l+n-1),\quad E(r)=\frac{Q\sqrt{n(n-1)}}{r^{n}}\,,\\
			&{\mathcal A}(r)=2\kappa_s+nrf^{\prime}-2nf\,,\\
			&T(r)=r^{n-1}\left[2\alpha(n-1)(n-2)\Psi(r)+1\right]\,,\\
			&\Psi(r)=\frac{1-f(r)}{r^2}\,.
		\end{split}
	\end{equation}
	We have also introduced the tortoise coordinate $r_*$ which is defined as $dr_* =dr/f$. In contrast to the Einstein-Maxwell theory~\cite{Kodama:2003kk} where one can find one master equation in the scalar type perturbations by manipulating the Einstein and Maxwell equations, for the case with GB term~\eqref{GB}, the gravito-electromagnetic perturbations are coupled with each other. More precisely, the variable $\phi_{\omega l}$ comes from gravitational perturbations, while $A_{\omega l}$ is obtained by a particular combination of gravitational and Maxwell perturbations. Here the subscript $\omega$ denotes the frequency measured in units of $r_c$, and $l=2, 3, 4,\dots$ is the multipole number. Modes with $l=0,1$ are pure gauge~\cite{Kodama:2003kk}. The instability region of the background black hole~\eqref{RN} will be obtained by solving the two coupled equations~\eqref{Meq} with relevant boundary conditions.

	\section{Instabilities from near-horizon criterion}\label{sec:BF}
	The two coupled master equations~\eqref{Meq} do not have analytic solutions, so one has to solve them numerically. There is an instability criterion conjectured by Durkee and Reall~\cite{Durkee:2010ea} which claims that when the near-horizon effective mass violates the $AdS_2$ BF bound, then the full spacetime geometry is linear unstable. This local criterion was proved later for asymptotically flat ($\Lambda=0$) and $AdS$ ($\Lambda<0$) black holes~\cite{Hollands:2014lra}. It was also argued to be valid in de Sitter case for which $\Lambda>0$. Recently, the authors of~\cite{Dias:2020ncd} performed a detailed analysis of the near-horizon limit of extremal RNdS black hole and provided good numerical evidence for the Durkee-Reall conjecture with $\Lambda>0$. Before doing numerics, we shall consider the near-horizon limit of the extremal charged black hole~\eqref{RN} and try to see if there exists any mode that violates the $AdS_2$ BF bound.

	\subsection{Near-horizon limit at extremality}
	The near-horizon geometry of the extremal GB-RNdS black hole~\eqref{RN} can be obtained as follows. One first takes $Q=Q_{ext}$, and then zooms in near the event horizon region by considering the following 
	coordinate transformation.
	\begin{equation}\label{IRlimit}
		\begin{split}
			r\rightarrow r_++\epsilon\rho,\quad t\rightarrow \frac{R_2^2}{\epsilon}\tau, \quad R_2^2=\frac{2}{f^{''}(r_+)}\,,
		\end{split}
	\end{equation}
	with $\epsilon$ a constant. Then, by taking the limit $\epsilon\rightarrow 0$, one obtains the near-horizon geometry
	\begin{align}\label{AdS2}
		ds^2&=R_2^2\left(-\rho^2 d\tau^2+\frac{1}{\rho^2}d\rho^2\right)+r_+^2d\Omega_n^2,\\
		A&=R_2^2A_t'(r_+)\rho d\tau\,,
	\end{align}
	which is the direct product of $AdS_2\times S^n$. We point out that the two master variables $\phi_{\omega l}$ and $\mathcal{A}_{\omega l}$ do not change under above operation.
	
	Taking the same near-horizon limit as~\eqref{IRlimit} together with $\hat{\omega}=\epsilon \omega/ R_2^2$ on the master equations~\eqref{Meq}, we obtain that 
	\begin{align}
		{\rho^2}\frac{\mathrm{d}^2}{\mathrm{d}\rho^2}
		\begin{bmatrix}
			\phi_{\omega l}\\
			A_{\omega l}
		\end{bmatrix}+{2\rho}\frac{\mathrm{d}}{\mathrm{d}\rho}\begin{bmatrix}
			\phi_{\omega l}\\
			A_{\omega l}
		\end{bmatrix}-R_2^2\mathcal{M}\begin{bmatrix}
			\phi_{\omega l}\\
			A_{\omega l}
		\end{bmatrix}=0\,,
	\end{align}
	with the mass matrix 
	\begin{align}
		\mathcal{M}=\begin{bmatrix}
			\frac{\kappa_s}{n r_+}\left(\frac{2(\kappa_s-n)}{\kappa_s}\frac{T'}{T}-\frac{T''}{T'}\right)& -\sqrt{\frac{\kappa_s-n}{n}}\frac{\sqrt{T'}}{T}\frac{2E}{r_+^{1-n/2}}\\
			-\sqrt{\frac{\kappa_s-n}{n}}\frac{\sqrt{T'}}{T}\frac{2E}{r_+^{1-n/2}}& \frac{\kappa_s}{r_+^2}\left(1+\frac{2E^2}{\kappa_s{T}}r_+^{n+1}\right)
		\end{bmatrix}\,,
	\end{align}
	evaluated at $r_+$ as well as $Q=Q_{ext}$. The eigenvalues of this matrix give the effective masses of the system. We denote its eigenvalues with $\mu^2_\pm$ and $\mu^2_-<\mu^2_+$. Therefore, one anticipates that there is an instability of the full charged black hole~\eqref{RN} once $\mu^2_-$ is smaller than the $AdS_2$ BF bound, \emph{i.e.} $\mu_{BF}^2 R_2^2=-1/4$.

	\begin{figure*}[ht]
		\includegraphics[width=0.45\textwidth]{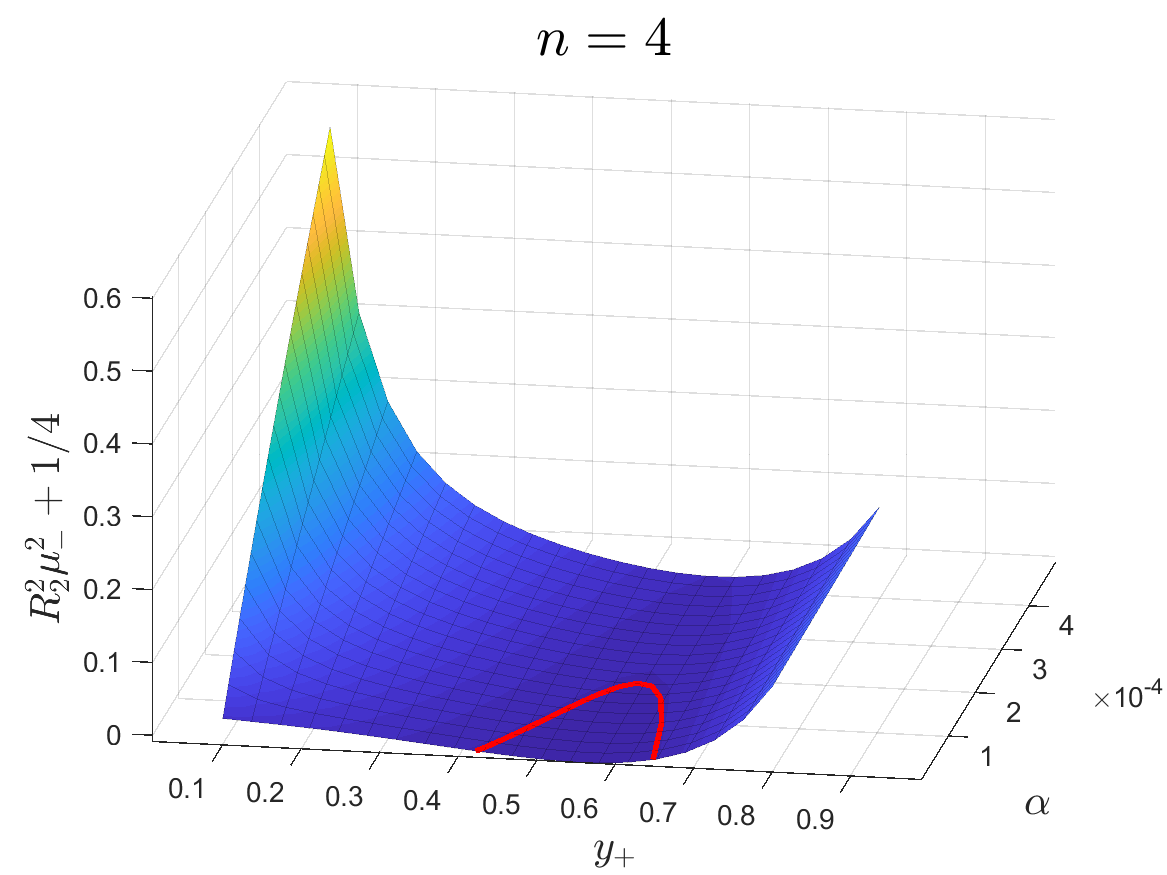}
		\includegraphics[width=0.45\textwidth]{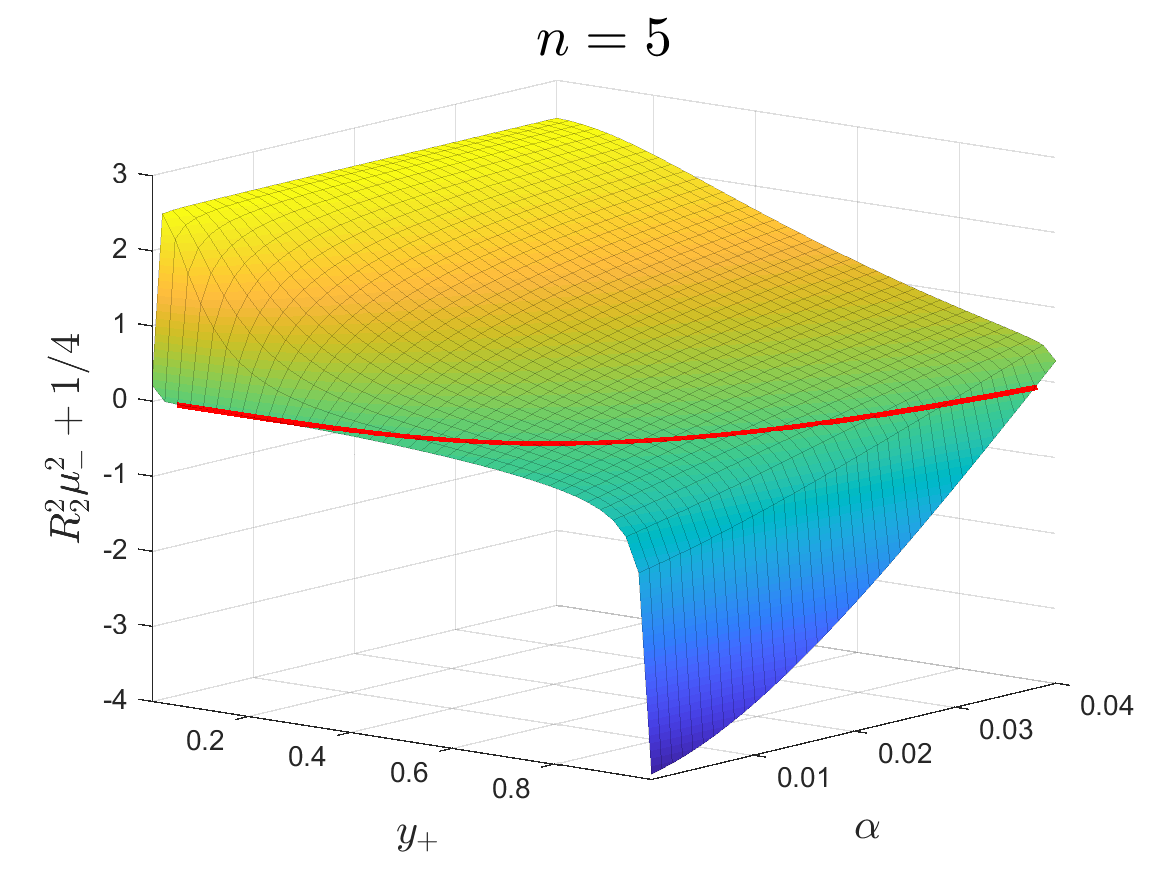}
		\caption{The effective near-horizon mass $\mu^2_-$ subtracted the $AdS_2$ BF bound $\mu^2_{BF}=-1/(4 R_2^2)$ as a function of $y_+$ and $\alpha$ for $n=4$ (left) and $n=5$ (right). The onset of the $AdS_2$ BF bound violation is marked by red curves.}
		\label{fig:BF3D}
	\end{figure*}
	
	\subsection{$AdS_2$ BF bound criterion}
	It has been recently shown that the effective mass of the scalar mode on the RNdS black hole~\eqref{BHsolution0} violates the BF bound when $n\ge4$~\cite{Dias:2020ncd}. We are interested in the effect of the GB term on the effective mass in the near-horizon region at extremality. As we have already mentioned, there are two independent parameters $\{y_+,\alpha\}$ describing the GB-RNdS black hole~\eqref{RN} at zero temperature. For each $n$, we will find the minimum of the effective mass and will check if it can violate the $AdS_2$ BF bound or not.  Except for $n=3$, we find that as the multipole number $l$ increases, it becomes more difficult to have a small value of the effective mass. Therefore, we shall mainly focus on the $l=2$ modes, but we will return to the $n=3$ case with higher $l$ in the next section. For illustration, we show $\mu^2_-$ as a function of $y_+$ and $\alpha$ for $n=4$ and $n=5$ in Fig.~\ref{fig:BF3D}.

\begin{figure*}[ht]
		\begin{center}
			\includegraphics[width=2.9in]{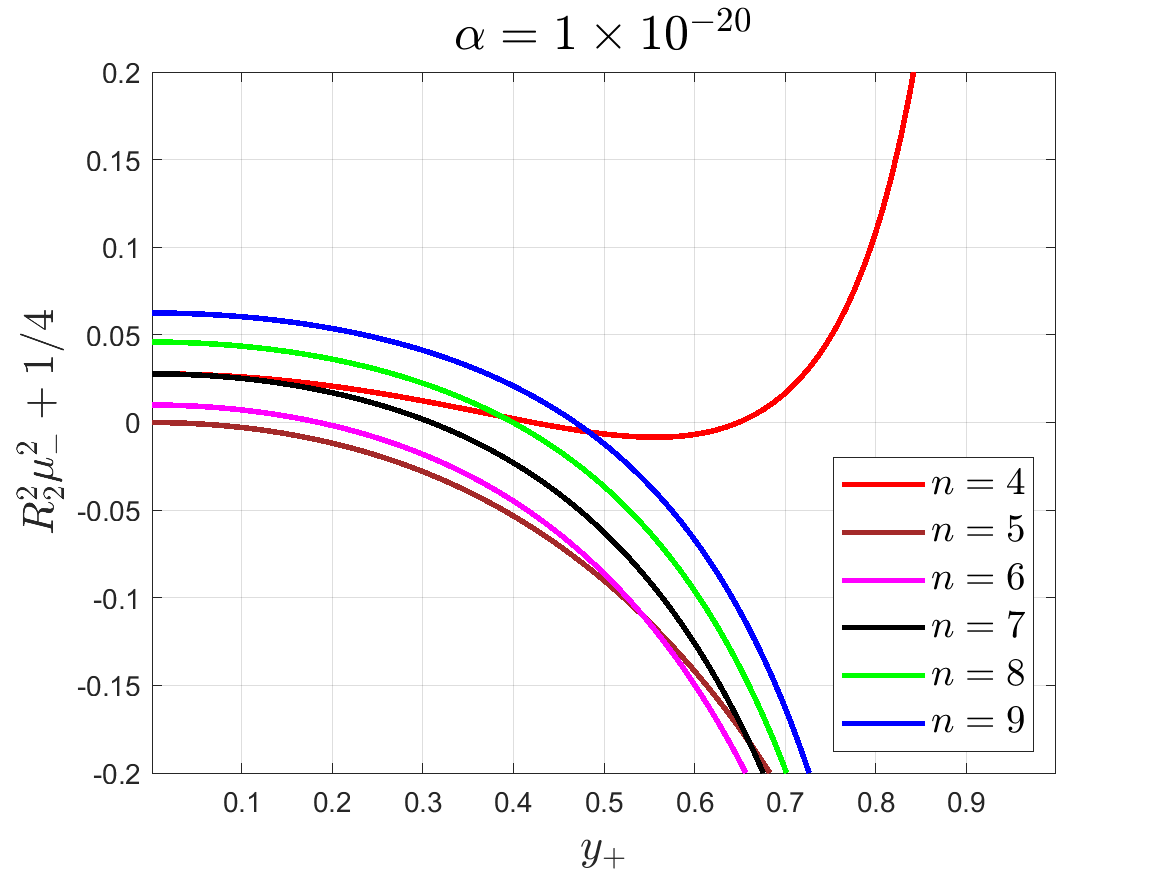}
			\includegraphics[width=2.9in]{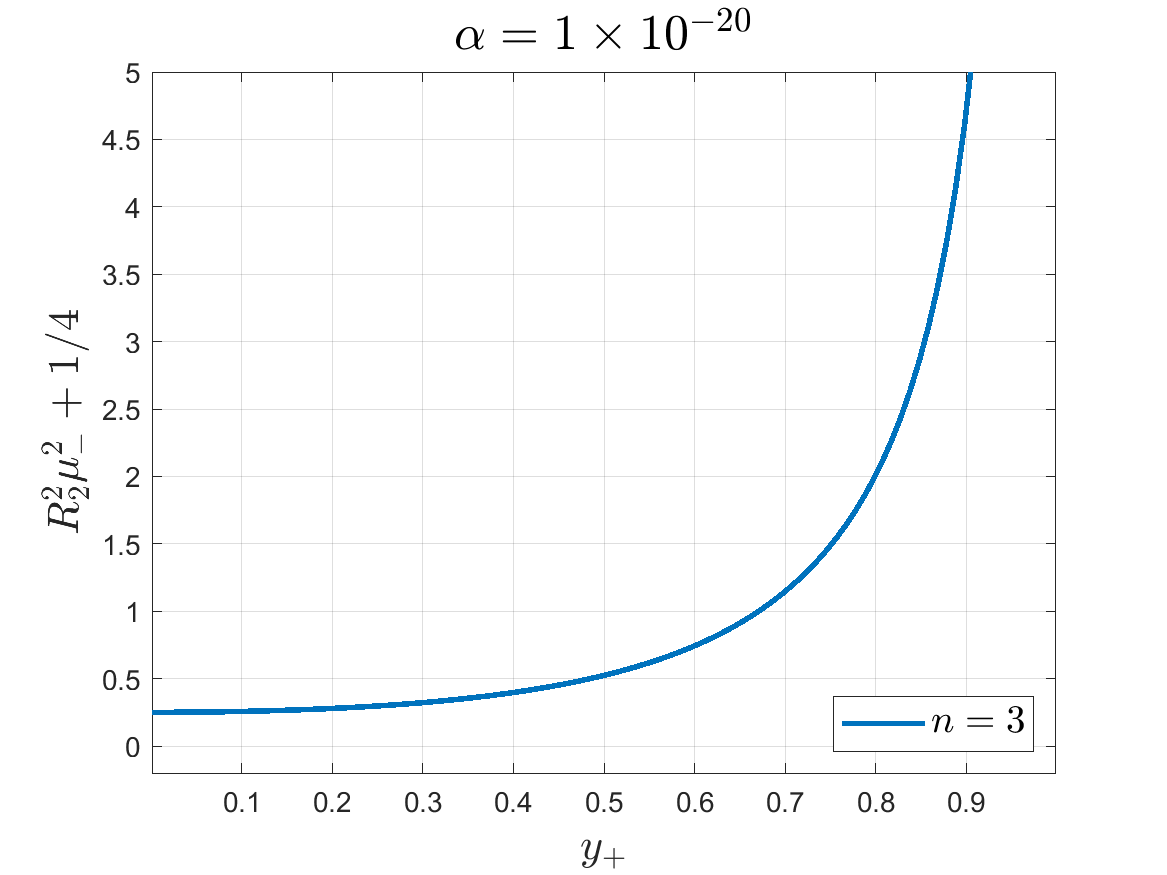}
			\caption{The effective near-horizon mass as a function of $y_+$ for the GB coefficient $\alpha=10^{-20}$. The BF bound  is violated when $\mu^2_-$ is smaller than $\mu^2_{BF}$.}
			\label{fig:alpha0}
		\end{center}
	\end{figure*}

	As a consistent check, let's consider the small $\alpha$ limit, for which one anticipates to recover the results of the RNdS black hole in~\cite{Dias:2020ncd}. In Fig.~\ref{fig:alpha0}, we take $\alpha=10^{-20}$ and plot $(\mu^2_--\mu_{BF}^2)R_2^2$ for different spacetime dimensions. We find that as $\alpha\rightarrow 0$, for $n\geqslant 4$ there always exist some values of $y_+$ for which the $AdS_2$ BF bound is violated, while there is no $AdS_2$ BF bound violation for $n=3$. In particular, we do recover the same result as Fig.~1 of~\cite{Dias:2020ncd} without the GB term by taking $\alpha\rightarrow0$ (see the left plot of Fig.~\ref{fig:alpha0}). This provides a consistent check for the master equations~\eqref{Meq} in Section~\ref{sec:setup}. 

\begin{figure*}[ht]
		\begin{center}
			\includegraphics[width=2.9in]{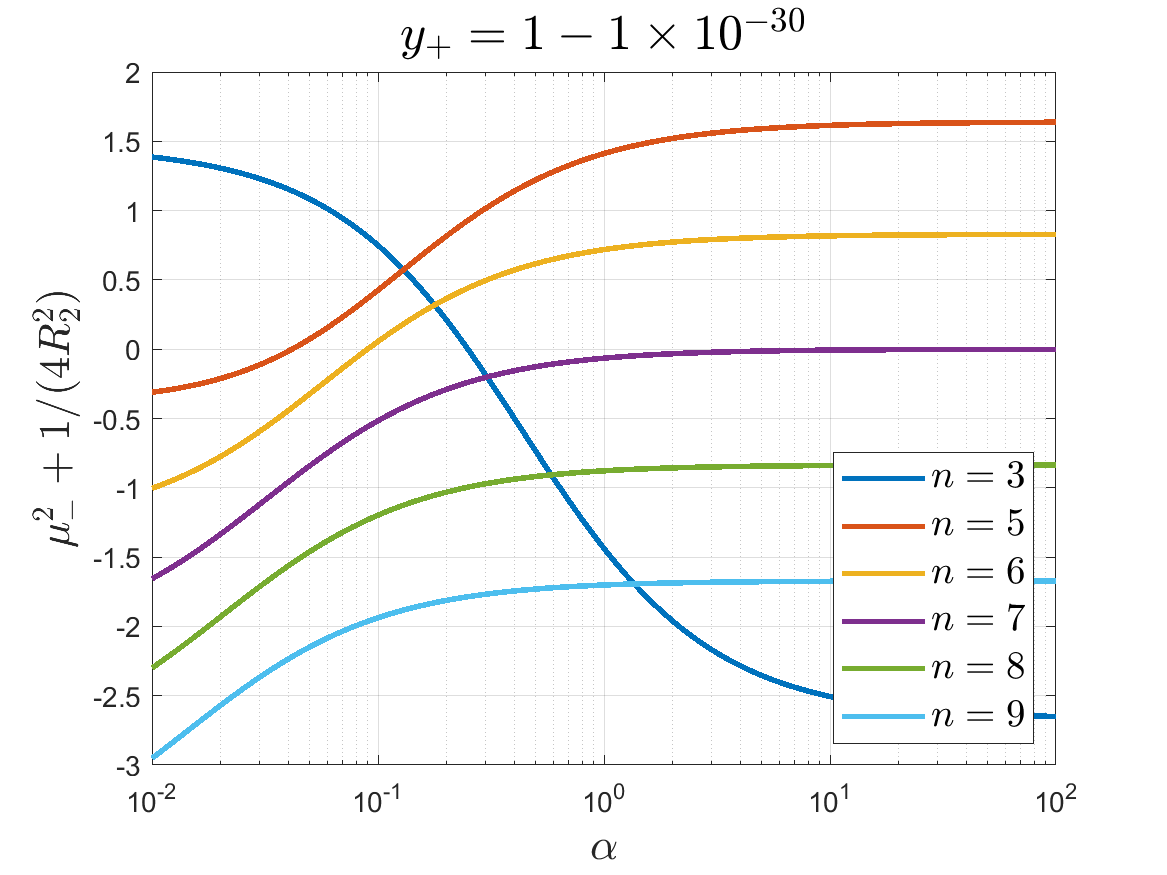}
			\includegraphics[width=2.9in]{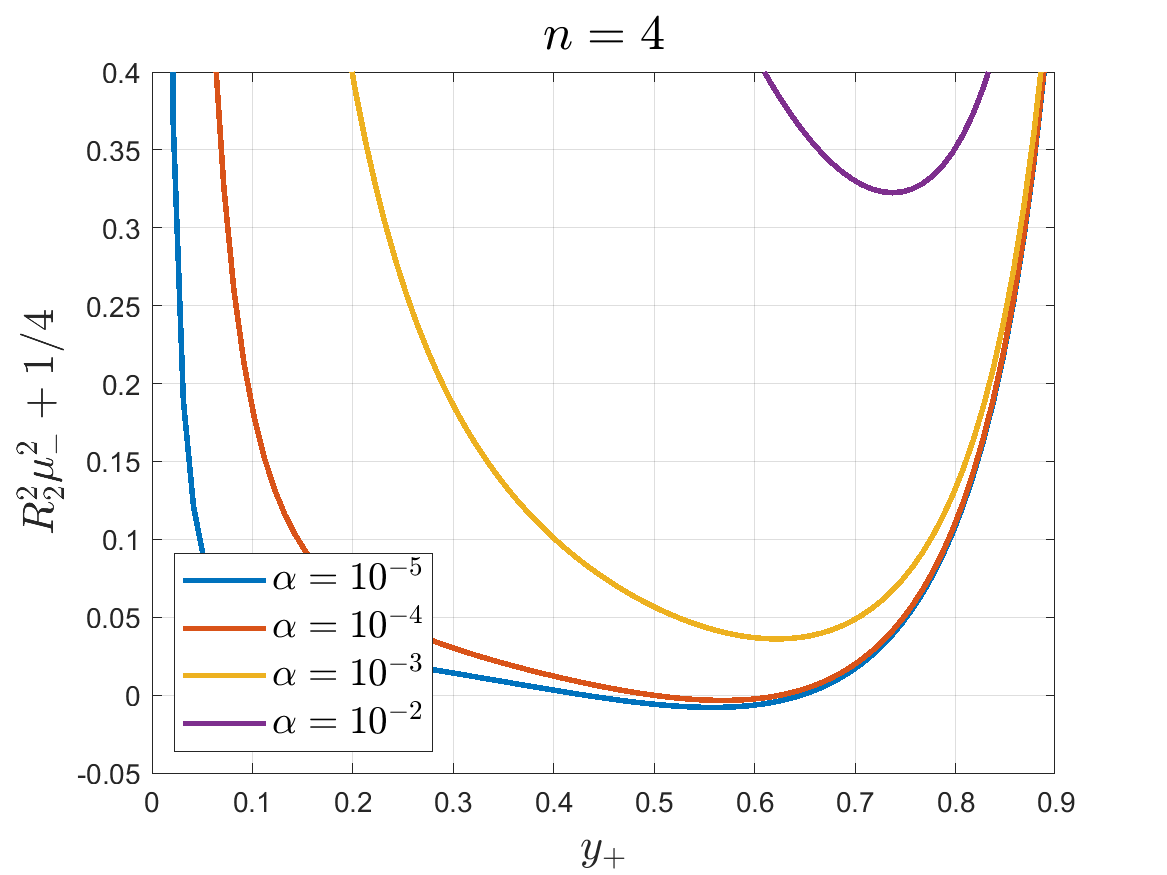}
			\caption{\textbf{Left panel}: $\mu_-^2-\mu^2_{BF}$ as function of $\alpha$ for $n=3,5,6,7,8$ at $y_+=1-10^{-30}$. \textbf{Right panel}: Effective mass as a function of $y_+$ for $n=4$ by dialing $\alpha$.}
			\label{fig:BF}
		\end{center}
	\end{figure*}

	We now increase the value of $\alpha$ to study the effect of $\alpha$ on the effective mass $\mu^2_-$. Except for $n=4$, for each $\alpha$ the minimum of $\mu^2_-$ is obtained as $y_+\rightarrow1$ (see Fig.~\ref{fig:BF3D}). Therefore, in the left panel of Fig.~\ref{fig:BF}, we show $\mu^2_--\mu_{BF}^2/R_2^2$ as a function of $\alpha$ for $n=3, 5, 6, 7, 8, 9$ with $y_+=1-10^{-30}$. For $n=4$, we plot $(\mu^2_--\mu_{BF}^2)R_2^2$ as a function of $y_+$ for different values of $\alpha$ in the right panel of Fig.~\ref{fig:BF}. We find three main features:
	\begin{itemize}
		\item For $n=3$, the effective mass $\mu^2_-$ is above the $AdS_2$ BF bound when $\alpha$ is small, but it decreases as $\alpha$ is increased and finally becomes lower than the $AdS_2$ BF bound.
		\item For $n=4,5,6$, the BF bound is violated for small $\alpha$, but the violation will disappear when $\alpha$ is sufficiently large. 
		\item For $n=7, 8, 9,\dots$, the value of $\mu^2_-$ increases as $\alpha$ is increased, but the $AdS_2$ BF bound violating modes always exist no matter how large $\alpha$ is.
	\end{itemize}

	Therefore, taking advantage of the Durkee-Reall instability criterion~\cite{Durkee:2010ea,Hollands:2014lra}, we are able to make some predictions from the local near-horizon analysis. While the RNdS black hole~\eqref{BHsolution0} is stable in five dimensions, we find that the GB-RNdS case would become unstable when $\alpha$ is large enough. In contrast, while the RNdS background is unstable for $n=4, 5, 6$, the GB-RNdS one seems to be stable for sufficiently large $\alpha$. For $n\geqslant 7$, there are always BF bound violating modes and an instability should be present. In the next section, we will compute the QNMs by numerically solving two master equations~\eqref{Meq}. We will compare our numerical results with the present near-horizon analysis.

	\section{Dynamical instabilities from QNMs}\label{sec:qnm}
	In this section we aim at the QNM spectrum of linear perturbations in the GB-RNdS black hole~\eqref{RN}. See~\cite{Konoplya:2011qq} for a recent review on the QNMs of black holes. More precisely, if the imaginary part of a QNM frequency, $\text{Im}[\omega]$, is positive, such mode will grow exponentially with time, signifying a dynamical instability. In particular, the value of a positive $\text{Im}[\omega]$ sets the timescale of the instability. To directly compare with the analytical near-horizon prediction in the last section, we shall mainly focus on the extremal case. 
	
	The QNMs are obtained by solving those master equations~\eqref{Meq} between the event and cosmological horizons with the ingoing boundary condition at the event horizon and the outgoing boundary condition at the cosmological horizon. In our present case, the master equations have a cumbersome form, which makes the numerical analysis of the QNM spectrum challenging. In particular, we find that the double-precision (16 digits) arithmetics are not sufficient for our computations. Therefore, higher precision is required when solving the master equations. Numerical techniques and details are presented in Appendix~\ref{app:detail}. 
	
	We now present the numeric results for the QNMs by solving two master equations~\eqref{Meq}. As we have discussed, a dynamical instability will develop if the imaginary part of a QNM frequency $\omega$ is positive. We shall mainly focus on the extremal black holes ($Q/Q_{ext}=1$) which are parametrized by $y_+$ and $\alpha$. In particular, we will compare the parameter space $\{y_+, \alpha\}$ of instability from the QNMs and the one from the local $AdS_2$ BF bound violation. {We search directly for the unstable modes in the non-extremal GB-RNdS black hole in subsection~\ref{sec:Tneq0}.} 
	
	\subsection{The case for $n=3$}
\begin{figure*}[ht]
		\begin{center}
			\includegraphics[width=2.9in]{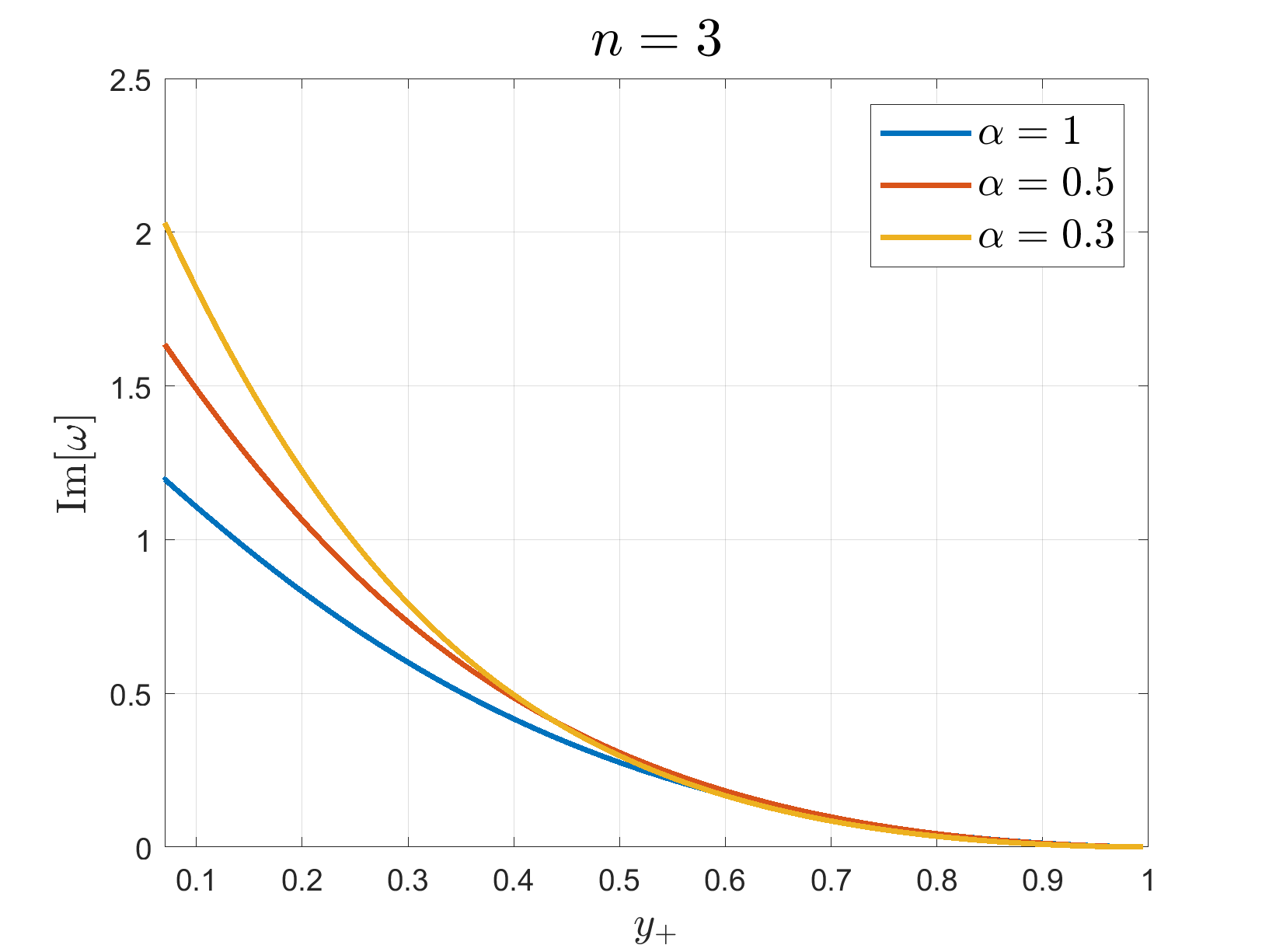}
			\includegraphics[width=2.9in]{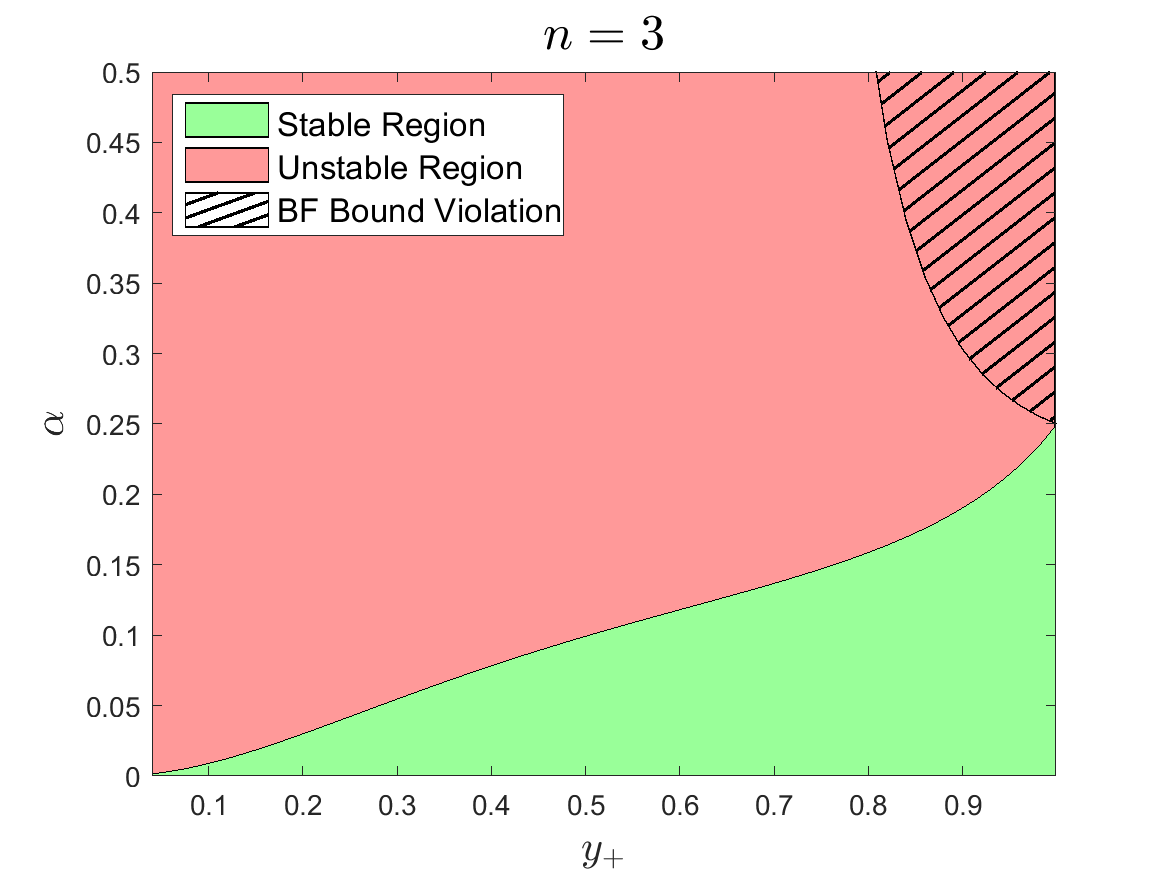}
			\caption{\textbf{Left panel}: The imaginary part of the QNM frequency $\text{Im}[\omega]$ as a function of $y_+$ for $n=3$ at extremality. The instability timescale decreases as $y_+$ is increased. \textbf{Right panel}: The parameter space for the  linear stable and unstable regions in the $y_+$-$\alpha$ plane for $n=3$ ($d=5$). We have focused on the lowest multipole number $l=2$.}
			\label{fig:region_n3}
		\end{center}
	\end{figure*}

	The main numerical results for $n=3$ are summarized in Fig.~\ref{fig:region_n3}. Depending on the choice of $y_+$ and $\alpha$, there exist QNMs with $\text{Im}[\omega]>0$ for which the background is dynamically unstable (see the left panel of Fig.~\ref{fig:region_n3}). The parameter space in the $y_+$-$\alpha$ plane at extremality is presented in the right panel of Fig.~\ref{fig:region_n3}. In the green region at the bottom right, we find that all QNM frequencies have negative imaginary part and thus no instabilities. For other regions, we can find unstable modes with $\text{Im}[\omega]>0$, signifying a dynamical instability of the background spacetime. The region that violates the $AdS_2$ BF bound is included entirely in the unstable region (red color) obtained from QNMs, suggesting that the BF bound violation is a sufficient but not necessary condition for the presence of an instability.

	We highlight the following three features:
	\begin{itemize}
		\item The unstable region shrinks as the GB coupling  $\alpha$ decreases, but it disappears only in the limit $\alpha\rightarrow 0$. Therefore, in contrast to the Einstein-Maxwell case for which the RNdS black hole~\eqref{BHsolution0} is stable for $n=3$  ($d=5$), the GB-RNdS black hole~\eqref{RN} {has some unstable region in parameter space} no matter how small $\alpha$ is (Note that $\alpha=\bar{\alpha}/r_c^2$. Our numerics shows that in the right panel of Fig.~\ref{fig:region_n3}, the critical line separating the stable and unstable regions behaves as $\alpha\sim y_+^2$ for small $y_+$. In unites of $\bar{r}_+$, one then finds that the instability will appear beyond a finite value of $\bar{\alpha}/r_+^2=\alpha/y_+^2$. We will return to this point in subsection~\ref{sec:highl}.). 
		
		\item  The $AdS_2$ BF bound violating modes appear for a large value $\alpha>\alpha_0\approx0.25$.  Although typically the near-horizon analysis predicts instability for a finite range of $y_+$, the system is unstable in the whole range $0<y_+<1$ at extremality (see the left panel of Fig.~\ref{fig:region_n3}). In this sense, the instability at extremality has a near-horizon origin, as argued  by~\cite{Dias:2020ncd} recently. However, for small $\alpha<\alpha_0$, there do not exist $AdS_2$ BF bound violation any more, but the instability is also present below a critical value of $y_+$  at a fixed $\alpha$.
		
		\item The unstable region from the QNMs and the one from BF bound violation coincide as $y_+\rightarrow 1$. This can be understood as follows. Note that $y_+=r_+/r_c$. When the event horizon is sufficiently close to the cosmological horizon, the whole region between $r_+$ and $r_c$ can be approximately described by the near-horizon $AdS_2$ geometry. Thus, one anticipates that the two criteria yield the same unstable region as $y_+\rightarrow 1$. 
	\end{itemize}

	We have to stress that the above results have been obtained for the lowest multipole number $l=2$. Nevertheless, it is sufficiently to show that the RNdS black hole with GB coupling in $d=5$ ($n=3$) dimensional spacetime is linearly-mode unstable, which is in contrast to the standard RNdS black hole which is linear stable in 5 dimensions~\cite{Kodama:2003kk}. As we will show later, for $n=3$ there is a dynamical instability occurring at high $l$. So the instability region in the 5 dimensions will be larger than the one presented in Fig.~\ref{fig:region_n3}, but the main features we discuss here will not change.
	
	\subsection{The case for $n=5,6$}\label{sec:n56}
	For $n=5$ and $n=6$, the near-horizon analysis shows that there is a range of $y_+$ in which the $AdS_2$ BF bound is violated when $\alpha$ is small. Then the Durkee-Reall criterion~\cite{Durkee:2010ea,Hollands:2014lra} predicts that the gravitational instability in the GB-RNdS black hole appears for these values of $\alpha$. When $\alpha$ is large enough, the BF bound violating modes will disappear, for which the Durkee-Reall criterion no longer applies and one has to switch to the full numerical analysis.

	\begin{figure}[ht]
		\centering
		\includegraphics[width=0.45\textwidth]{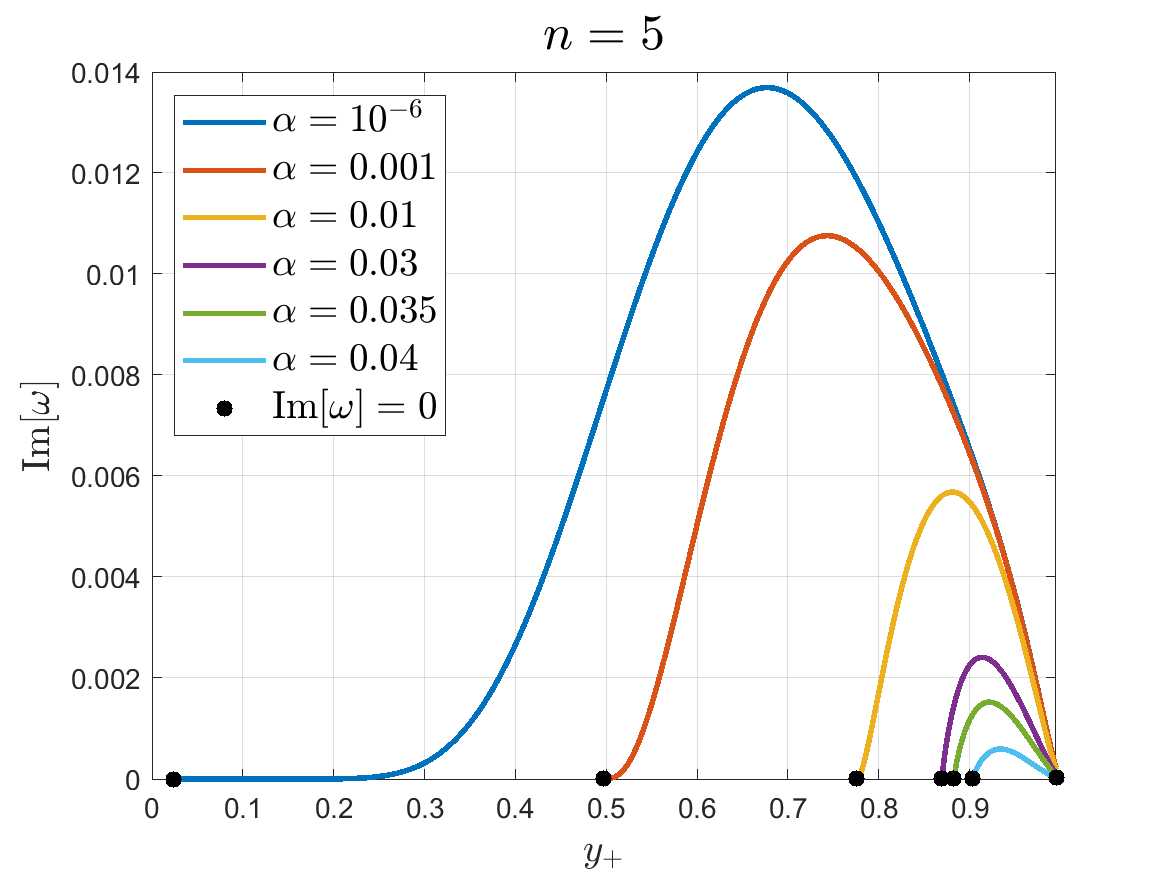}
		\caption{Instability timescale at extremality for $n=5$ as a function of $y_+$ for different values of $\alpha$. The onset of instability at which \text{Im}$[\omega]=0$ is marked by black dots.}\label{fig:alpha_n5}
	\end{figure}
	In Fig.~\ref{fig:alpha_n5}, we show the behavior of $\text{Im}[\omega]$ for the unstable modes as a function of $y_+$ for $n=5$ by dialing $\alpha$. As shown in the left panel of Fig.~\ref{fig:BF}, for $n=5$, as $\alpha$ increases it becomes more difficult to get negative $(\mu_-^2-\mu^2_{BF})R_2^2$, and the $AdS_2$ BF bound is no longer violated when $\alpha$ is large enough. Our numerics analysis confirms that when the instability is present, modes with smaller $\alpha$ are more unstable. As one can see from Fig.~\ref{fig:alpha_n5}, the magnitude of the instability timescale decreases as $\alpha$ is increased. In contrast to $n=3$, for a given $\alpha$ that allows a finite range of $y_+$ with BF bound violation, the instability does not present in the full range $0<y_+<1$. Moreover, we find that the unstable region of $y_+$ shrinks by increasing $\alpha$. Similar behavior is also found for $n=6$.

\begin{figure*}[ht]
		\begin{center}
			\includegraphics[width=2.9in]{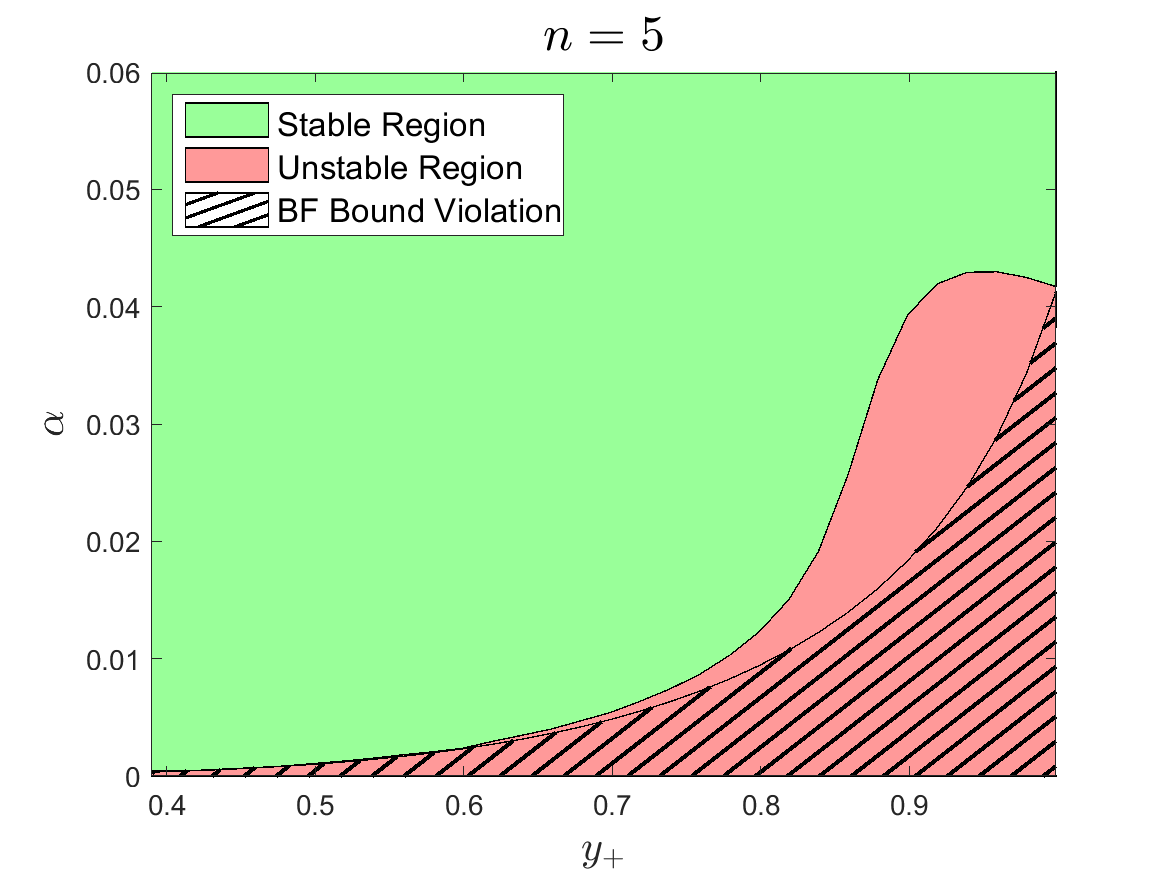}
			\includegraphics[width=2.9in]{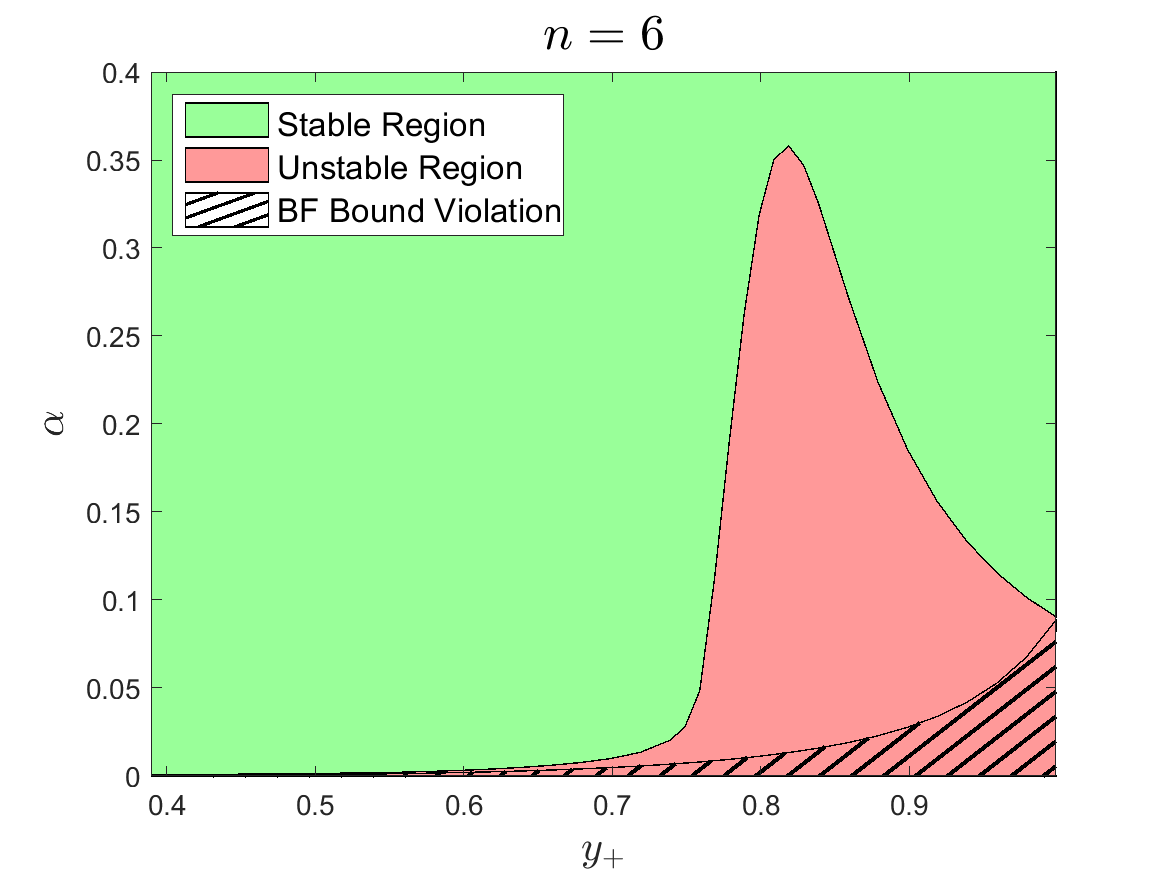}
			\caption{The parameter space at extremality for the linear stable and unstable regions in the $y_+$-$\alpha$ plane for $n=5$ (\textbf{left}) and $n=6$ (\textbf{right}). The instability disappears for sufficiently large GB coupling.}
			\label{fig:region_n56}
		\end{center}
	\end{figure*}

	The parameter space in the $y_+$-$\alpha$ plane at extremality is presented in Fig.~\ref{fig:region_n56} for $n=5$ (left plot) and $n=6$ (right plot). One can find that the system is indeed always unstable whenever the $AdS_2$ BF bound is violated. It has been observed that for the extremal RNdS black hole with $n\geqslant 4$, if the near-horizon analysis predicts instability for a finite range of $y_+$, the system will be unstable in the whole range $0<y_+<1$~\cite{Dias:2020ncd}. This feature is no longer valid for the GB-RNdS black hole. It is clear from Fig.~\ref{fig:region_n56} that although the instability extends to values of parameter space where the near-horizon effective mass does not violate the BF bound, it can not extend to the full range of $y_+$. See Appendix~\ref{app:detail} for numerical details we used to fix the boundary of the unstable modes.

	Finally, the near-horizon analysis shows that the BF bound violating modes disappear when $\alpha$ is large enough, say $\alpha>\alpha_{BF}$, while the full numerics finds that there is no instability above $\alpha_c$ that is larger than the one predicted from the near-horizon analysis. Therefore, the appearance of dynamical instabilities between these two GB couplings ($\alpha_{BF}<\alpha<\alpha_c$) does not have a locally near-horizon/extremal origin. 
	
	Once again, we confirm that the near-horizon $AdS_2$ BF bound violation is a sufficient but not necessary condition for the presence of an instability. On the other hand, for $n=5, 6$ ($d=7, 8$), the full numerical analysis suggests that the BF bound violation is not able to account for the physical origin of the instability for some range of GB coupling, \emph{i.e.} $\alpha_{BF}<\alpha<\alpha_c$ for which the $AdS_2$ BF bound is no longer violated for any $y_+$.

	\subsection{The case for $n\geqslant 7$}
	In this subsection we consider $n\geqslant 7$. As shown in Section~\ref{sec:BF}, the BF bound violating modes always exist no matter how large $\alpha$ is. Therefore, the near-horizon Durkee-Reall criterion~\cite{Durkee:2010ea,Hollands:2014lra} suggests that the gravitational instability in the GB-RNdS black hole is present for any value of the GB coupling $\alpha$ when $n\geqslant 7$. Furthermore, the near-horizon analysis also shows that as $\alpha$ increases, it becomes more difficult to get a smaller effective mass. Therefore, one anticipates that modes with smaller $\alpha$ are more unstable. These near-horizon predictions are confirmed by our numerical data. We find similar features for different $n\geqslant 7$. For illustration purposes below, we shall focus on $n=8$.

\begin{figure*}[ht]
		\begin{center}
			\includegraphics[width=2.9in]{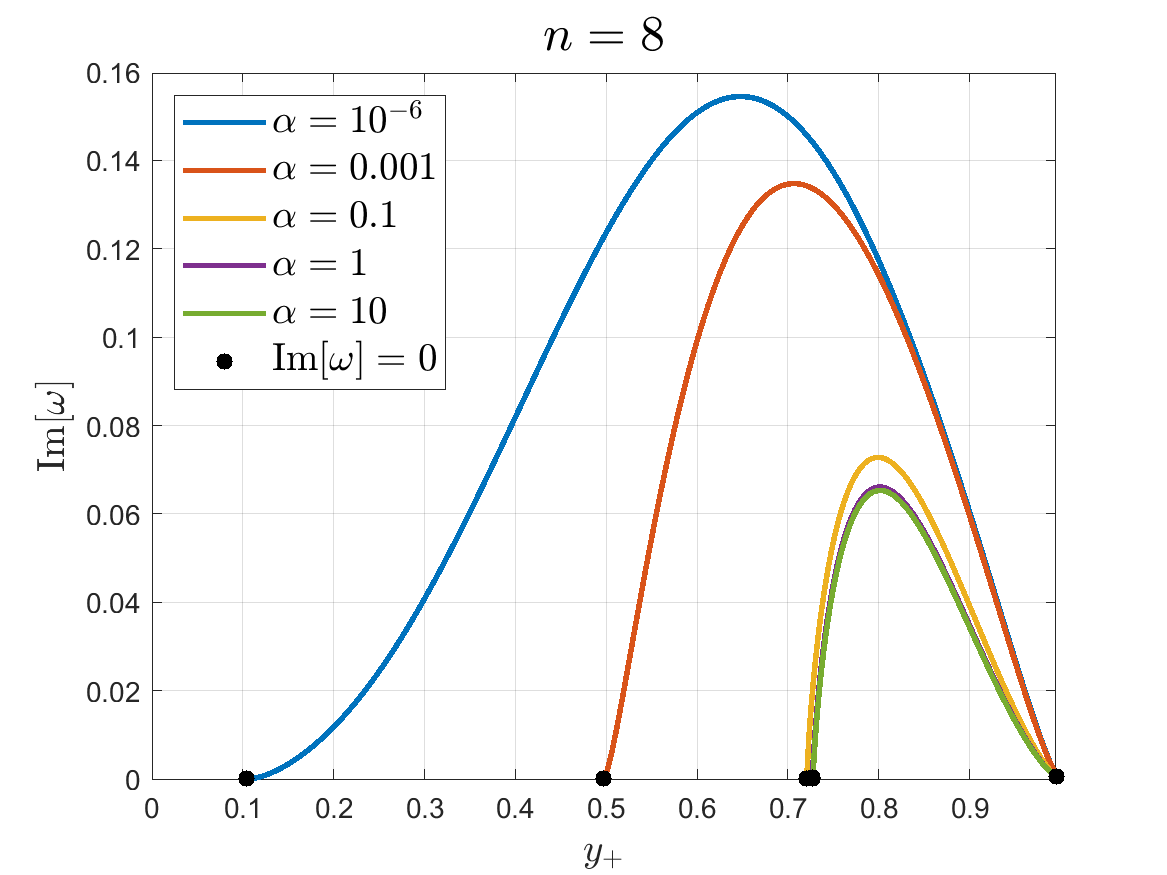}
			\includegraphics[width=2.9in]{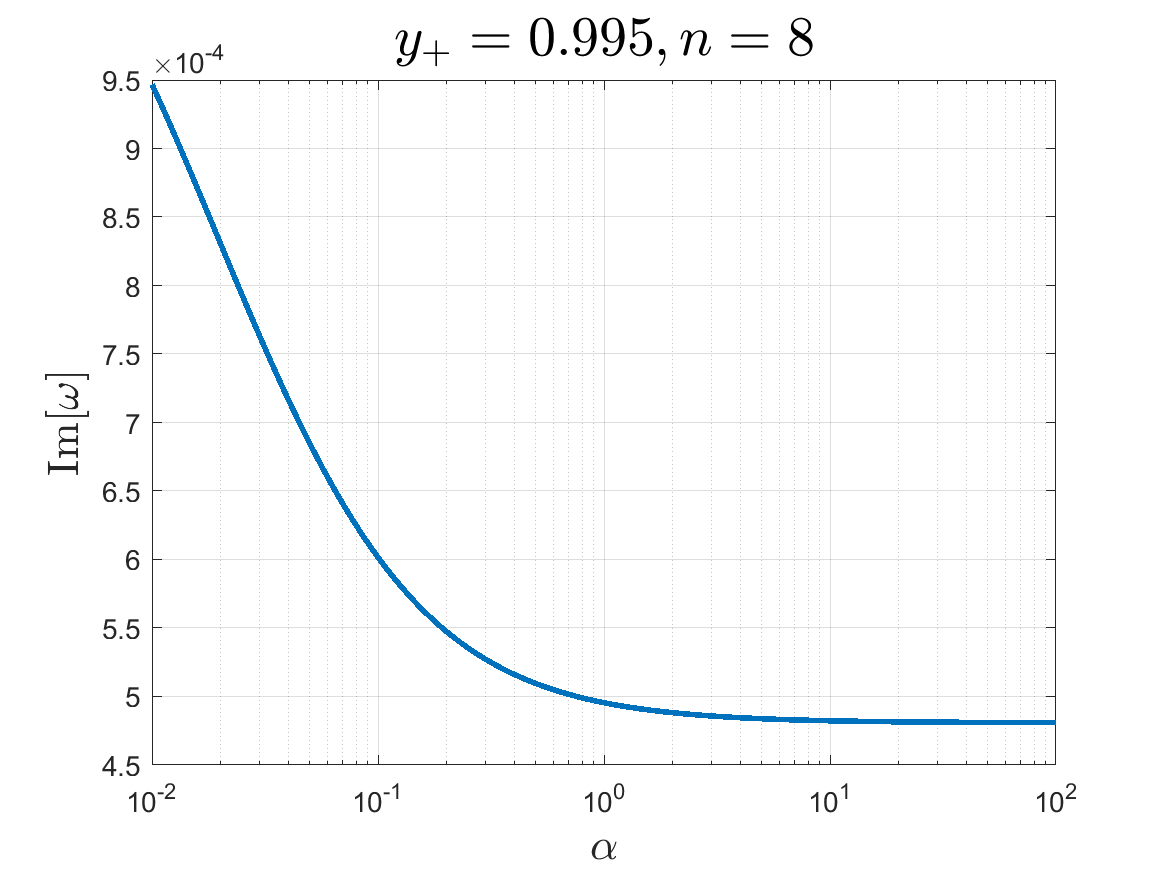}
			\caption{ Instability timescale at extremality for $n=8$. \textbf{Left panel}: Instability timescale as a function of $y_+$ for different $\alpha$. The threshold point of instability for which Im$[\omega]=0$ is marked by black dots. The instability does not extend to the full range of $y_+$ for the GB-RNdS black hole. \textbf{Right panel}: Instability timescale as a function of $\alpha$ by fixing $y_+=0.995$. The dynamcial instability is always present for any $\alpha$, but it becomes weaker at larger $\alpha$. }
			\label{fig:QNM_n8}
		\end{center}
	\end{figure*}

	In Fig.~\ref{fig:QNM_n8}, we compare the instability timescale of various unstable modes by dialing $\alpha$ for $n=8$. One can see that the magnitude of $\text{Im}[\omega]$ decreases as $\alpha$ is increased. Similar to the case for $n=5, 6$, the instability does not extend to the full range of $y_+$. More precisely, we find that the range of $y_+$ for the instability shrinks by increasing $\alpha$. We also explicitly check that the instability always exists for any value of $\alpha$. As one can see from the right panel of Fig.~\ref{fig:QNM_n8}, at a fixed $y_+$, $\mathrm{Im}[\omega]$ decreases monotonically and approaches to a positive constant with $\alpha$ increased.

\begin{figure}[ht]
		\centering
		\includegraphics[width=0.45\textwidth]{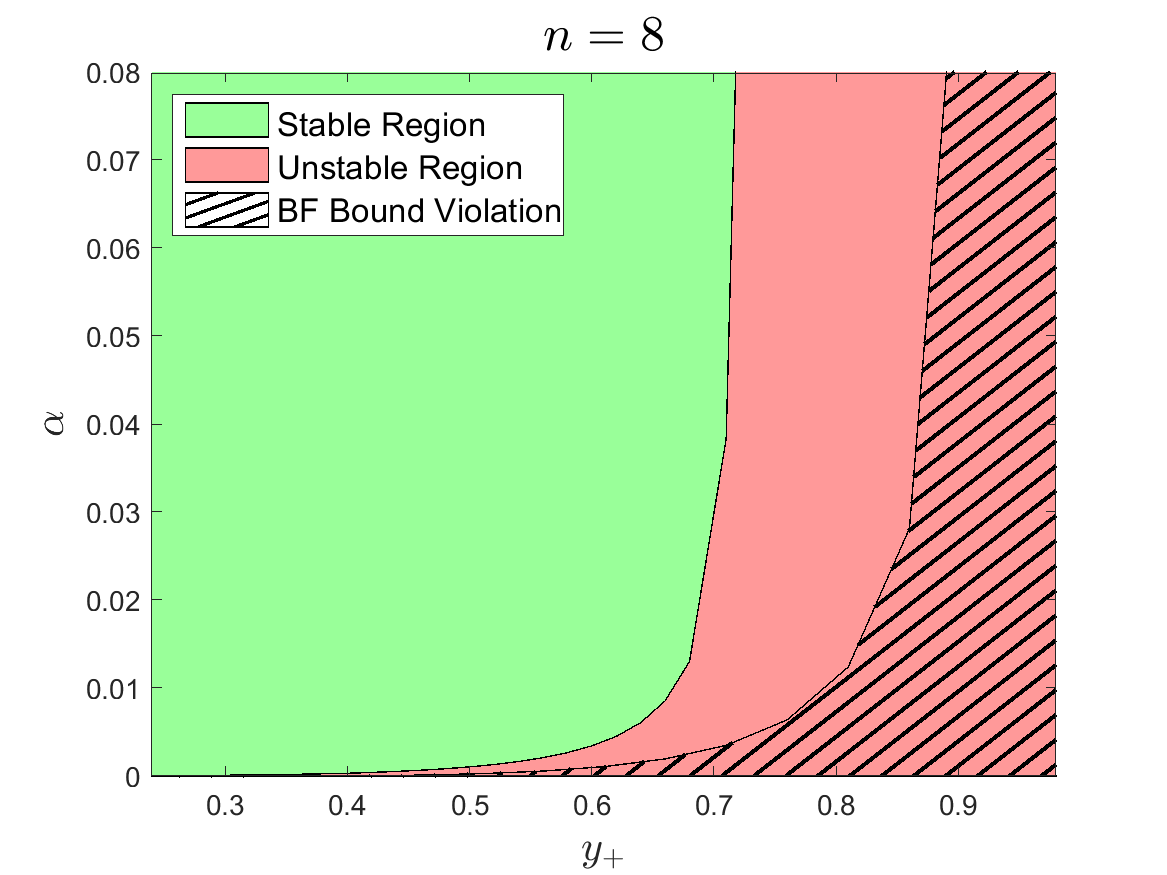}		
		\caption{Stability and instability regions for $n=8$ at extremality. The dynamical instability (red region) always exists for any value of the GB coefficient $\alpha$, but it does not extend to the full range of $y_+$.}\label{fig:region_n8}
	\end{figure}

	We draw the parameter space in the $y_+$-$\alpha$ plane at extremality in Fig.~\ref{fig:region_n8}. It is clear that the GB-RNdS black hole is always unstable whenever the $AdS_2$ BF bound is violated. On the other hand, although the instability extends to parameter space where the $AdS_2$ BF bound is no longer violated, it does not extend to the full range of $y_+$. Therefore, the violation of the $AdS_2$ BF bound is not able to explain the origin of the dynamical instability in the green region of Fig.~\ref{fig:region_n8}. Once again, we confirm that the Durkee-Reall criterion~\cite{Durkee:2010ea,Hollands:2014lra} provides a sufficient but not necessary condition for the presence of an instability.
	%

	\subsection{The case for $n=4$}
\begin{figure*}[ht]
		\begin{center}
			\includegraphics[width=2.9in]{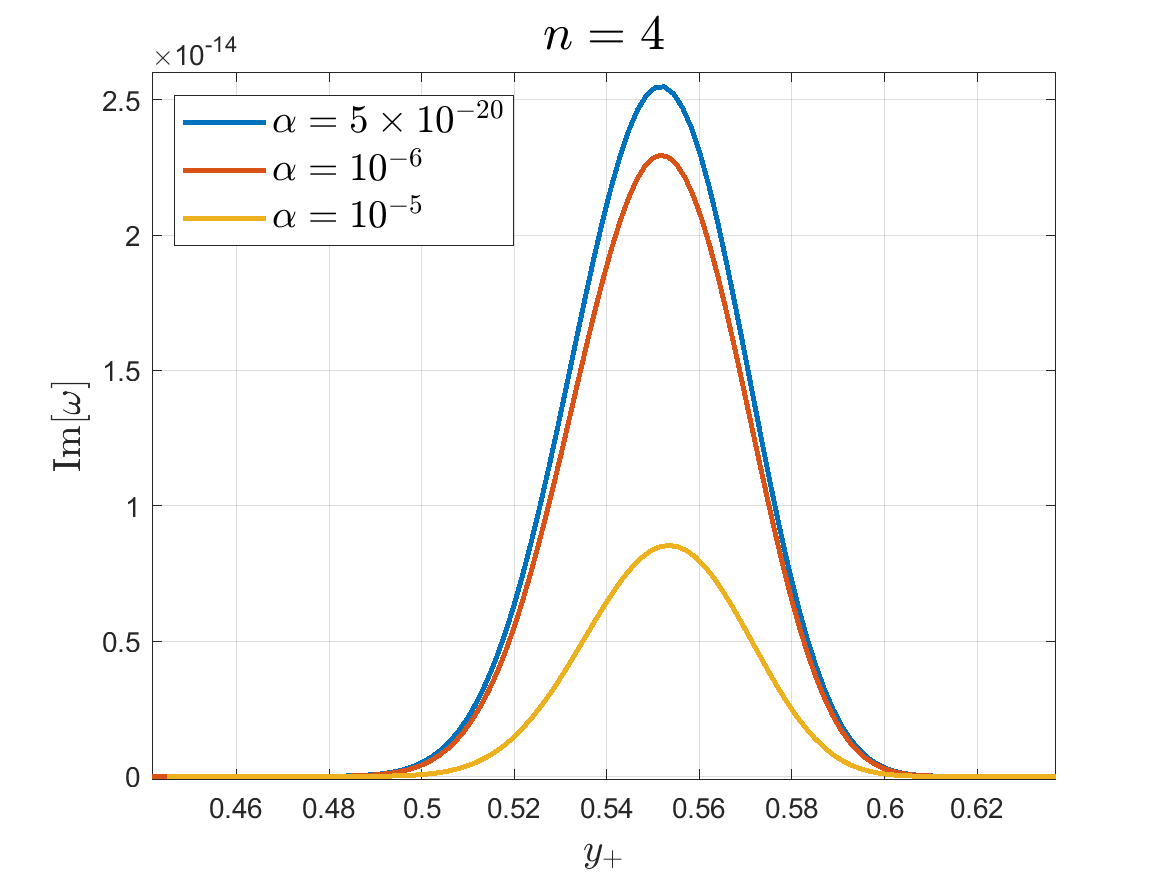}
			\includegraphics[width=2.9in]{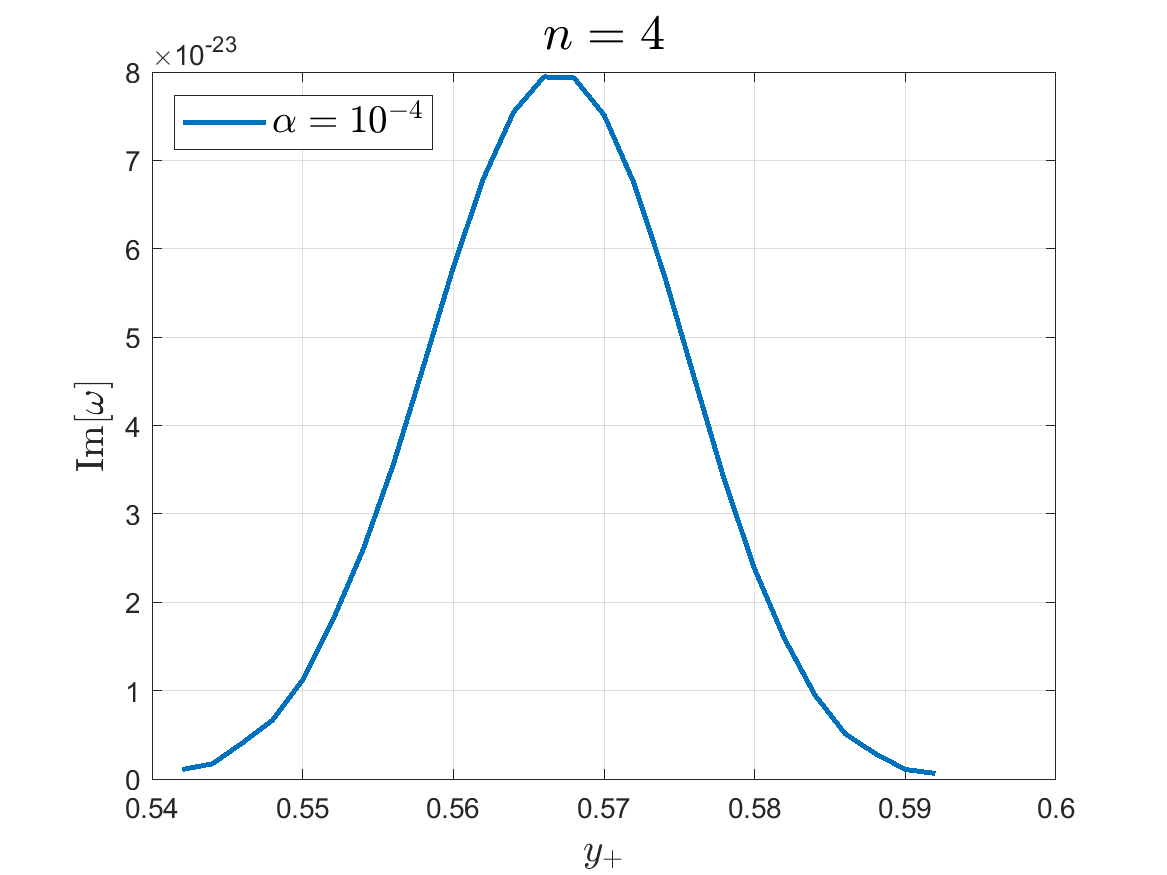}
			\caption{Instability timescale at extremality for $n=4$. \textbf{Left panel}: $\text{Im}[\omega]$ as a function of $y_+$ for different $\alpha$. The blue curve with $\alpha=5\times10^{-20}$ recovers the QNMs of the six dimensional RNdS black hole (the left panel of Fig.~2 in~\cite{Dias:2020ncd}). \textbf{Right panel}: The case with $\alpha=10^{-4}$, for which $\mathrm{Im}[\omega]$ is of the order of magnitude of  $10^{-23}$. }
			\label{fig:QNM_n4}
		\end{center}
	\end{figure*}

	So far we have discussed all cases except for $n=4$. For $n=4$, \emph{i.e.} $d=6$, the numerical analysis is challenging, because the instability growth rates are very small. It has been recently found in~\cite{Dias:2020ncd} that the gravitational instability in the RNdS black hole~\eqref{BHsolution0} is present when $n=4$, which is consistent with the prediction by the local near-horizon criterion. The RNdS background can be obtained from the GB-RNdS black hole~\eqref{RN} by taking the limit $\alpha\rightarrow 0$. Therefore, by continuity, the instability should extend to non-vanishing $\alpha$. Our numerical results are presented in Fig.~\ref{fig:QNM_n4} from which we indeed observe the dynamical instability for small values of $\alpha$. In particular, when $\alpha$ is sufficiently small, we recover the result of $\text{Im}[\omega]$ for the RNdS in six dimensions (the left panel of Fig.~2 in~\cite{Dias:2020ncd}), providing a non-trivial check of our numerics. 

\begin{figure}[ht]
		\centering
		\includegraphics[width=0.45\textwidth]{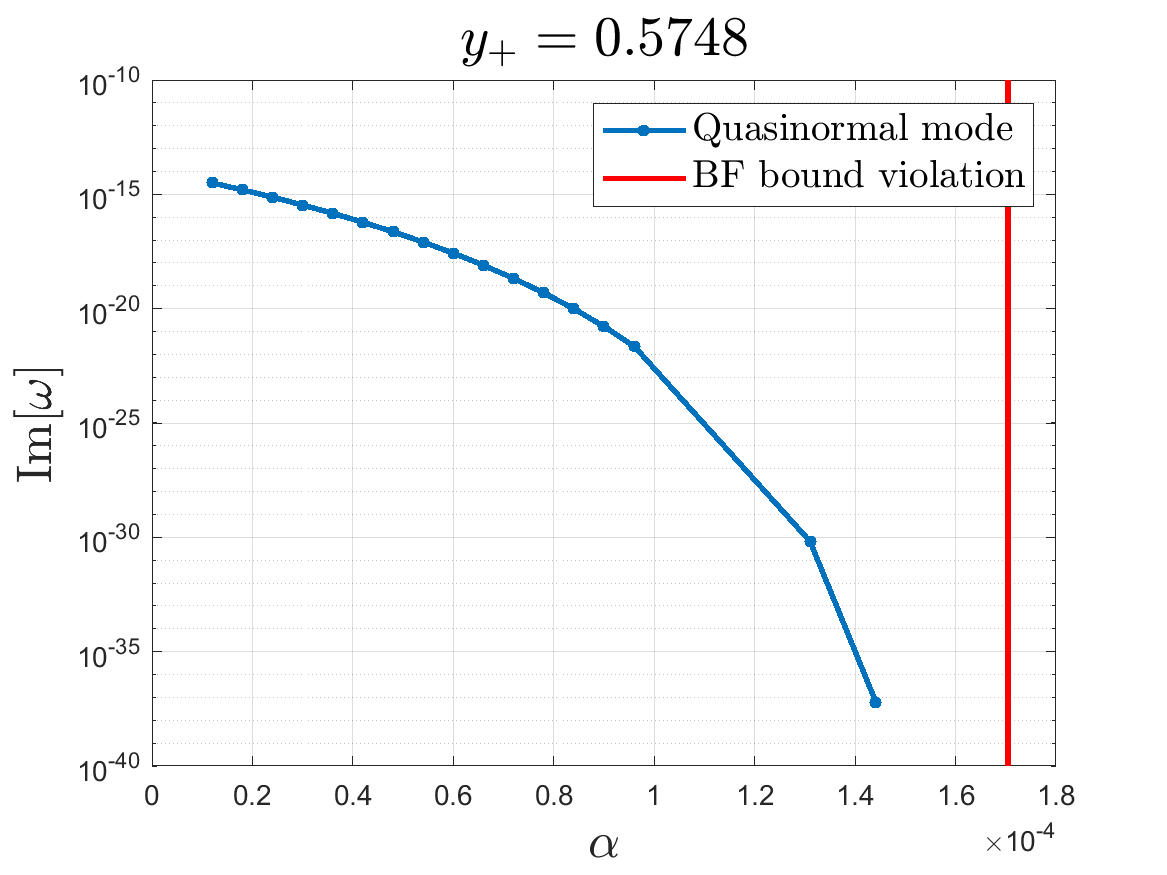}	
		\caption{Instability timescale at extremality for $n=4$. $\text{Im}[\omega]$ as a function of $\alpha$ at $y_+=0.5748$. The $AdS_2$ BF bound is no longer violated for $\alpha$ larger than the particular value denoted by the red vertical line.}\label{fig:critical_n4}
	\end{figure}

	The near-horizon analysis suggests that for $n=4$, the $AdS_2$ BF bound will not be violated when $\alpha$ is large enough, and therefore the instability would disappear for a sufficiently large value of $\alpha$. As one can see from Fig~\ref{fig:QNM_n4}, the instability indeed becomes substantially weaker as $\alpha$ is increased. Unfortunately, it is computationally challenging to generate the numerical data for $n=4$, and thus we are not able to know when the instability will eventually shut down as $\alpha$ is increased. In Fig~\ref{fig:critical_n4} the magnitude of Im$[\omega]$ drops substantially fast as $\alpha$ is increased and its value is found to be as low as $10^{-37}$ from our numerics subject to limited computational resources. Nevertheless, we believe this case should behave much like the $n=5, 6$ case in subsection~\ref{sec:n56}.

	\subsection{Eikonal instability at high multipoles}\label{sec:highl}
\begin{figure*}[ht]
		\begin{center}
			\includegraphics[width=2.9in]{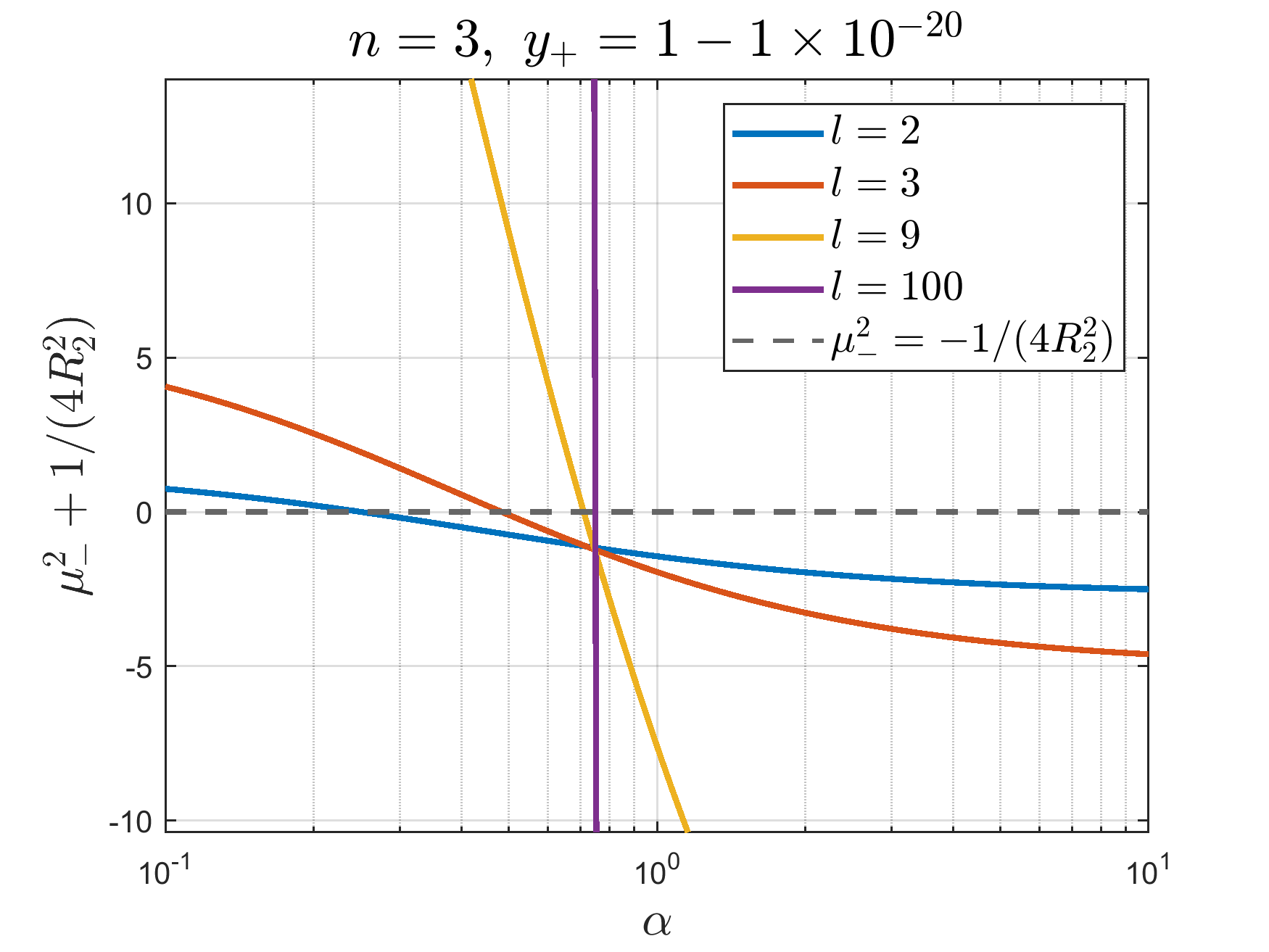}
			\includegraphics[width=2.9in]{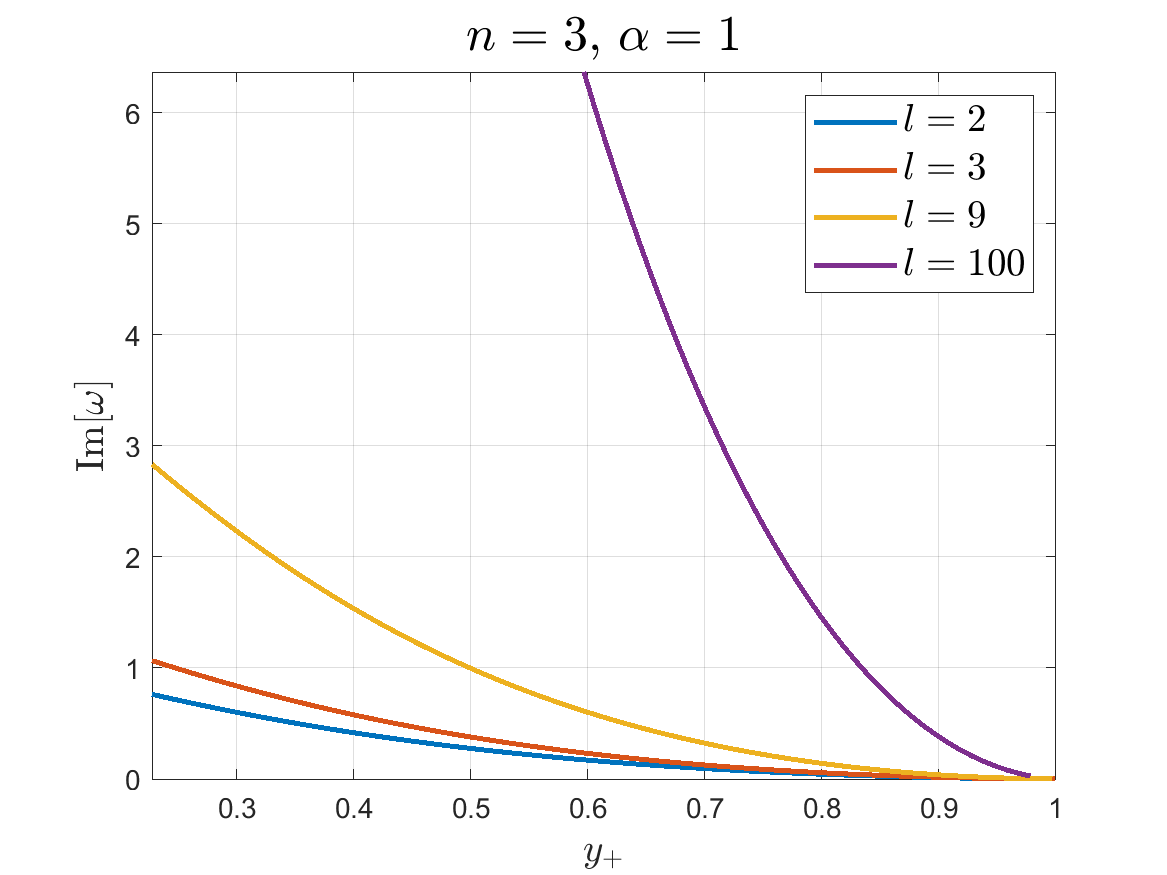}
			\caption{Instabilities triggered by higher multipole number $l$ for $n=3$ at extremality.  \textbf{Left panel}: BF bound violation at higher $l$ for a fixed $y_+$.  \textbf{Right panel}: Instability timescale as a function of $y_+$ for different $l$.  The higher $l$ modes are more unstable.}
			\label{fig:highl}
		\end{center}
	\end{figure*}

	In the above discussions, we have restricted to modes with $l=2$. This is because that for $n\geqslant 4$ the near horizon analysis shows that as $l$ increases it becomes harder to violate the $AdS_2$ BF bound. We have also confirmed from our numerics that the strength of instability for higher $l$ is much smaller than the $l=2$ instability. Nevertheless, the case is quite different from $n=3$, for which the BF bound becomes easier to be broken for higher $l$ modes, see the left panel of Fig~\ref{fig:highl}. We show the instability timescale of different modes in the right panel of Fig~\ref{fig:highl}. One finds that the instability timescale increases at higher $l$, confirming the analytical prediction from the BF bound point of view. 
	
	Therefore, in contrast to the instability that occurs for the lowest multipole $l=2$ we discussed before, the present instability develops at higher $l$ multipoles.  This kind of instability was called ``eikonal instability"~\cite{Cuyubamba:2016cug} to emphasize that the regime of geometrical optics ($l\rightarrow\infty$) corresponds to the most unstable mode. The eikonal instability was first observed in the static vacuum GB black holes in de Sitter spacetime~\cite{Cuyubamba:2016cug}, for which the origin of this instability is due to a negative gap of the effective potential near the event horizon. For the neutral case with GB term, each type of modes is governed by a decoupled master equation with a particular effective potential. Due to the GB correction, the effective potential develops a negative gap near the event horizon for some cases. Although increasing of $l$ would lead to a higher barrier of the effective potential, it also makes the negative gap deeper. Therefore, the modes with higher $l$ are more unstable~\cite{Cuyubamba:2016cug}. For the charged case we are discussing, we are not able to define an effective potential as there are two coupled master equations~\eqref{Meq}. Nevertheless, in contrast to the neutral case, the near horizon limit of the extremal charged black hole~\eqref{RN} is described by $AdS_2\times S^n$. As shown in the left panel of Fig~\ref{fig:highl}, the unstable modes with higher $l$ essentially break the $AdS_2$ BF bound. In particular, the higher $l$ is, the stronger the BF bound violation. Thus, we anticipate that the physical origin of the eikonal instability in the charged case is related to the violation of the $AdS_2$ BF bound.

\begin{figure*}[ht]
		\begin{center}
			\includegraphics[width=2.9in]{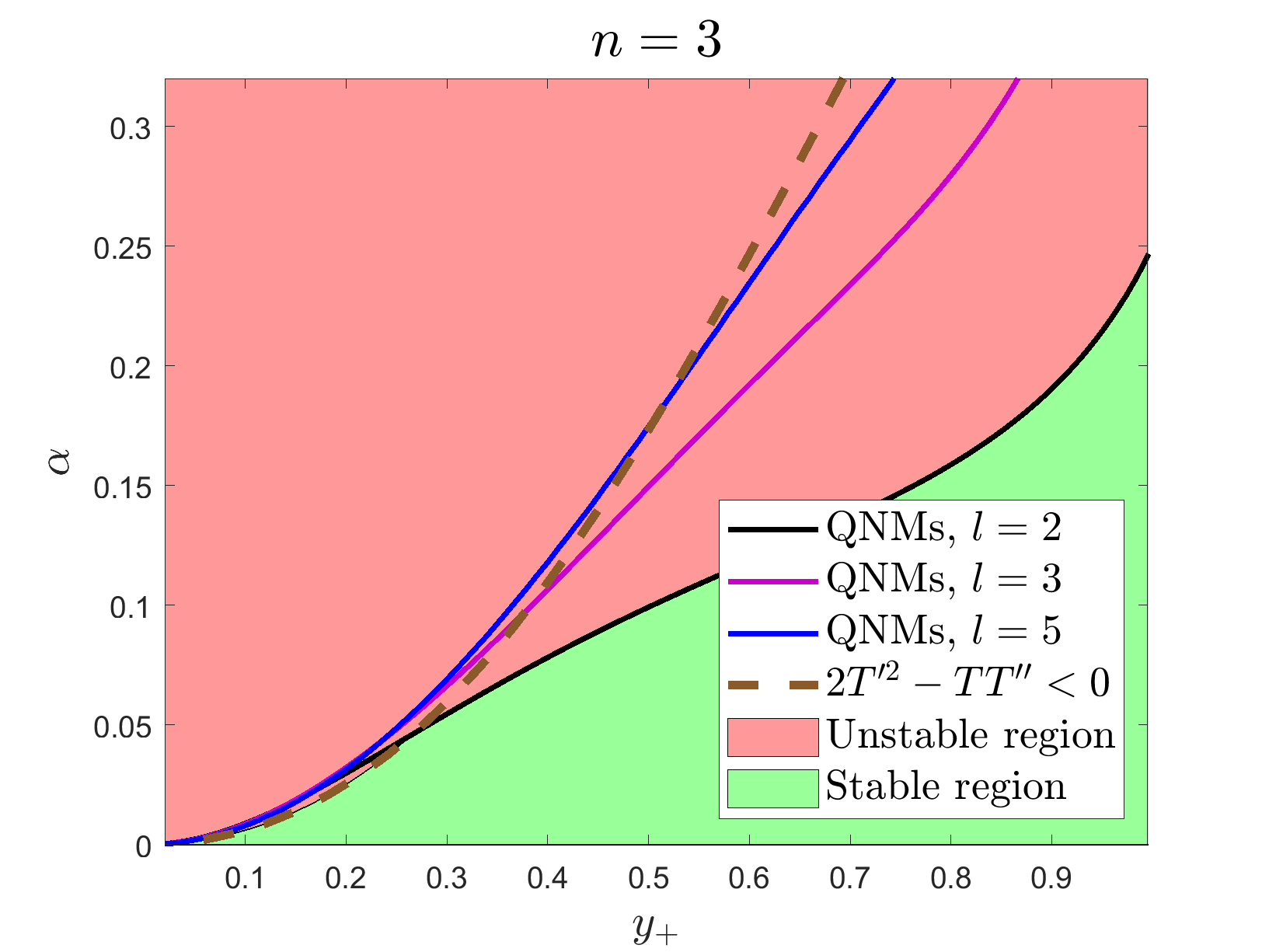}
			\includegraphics[width=2.9in]{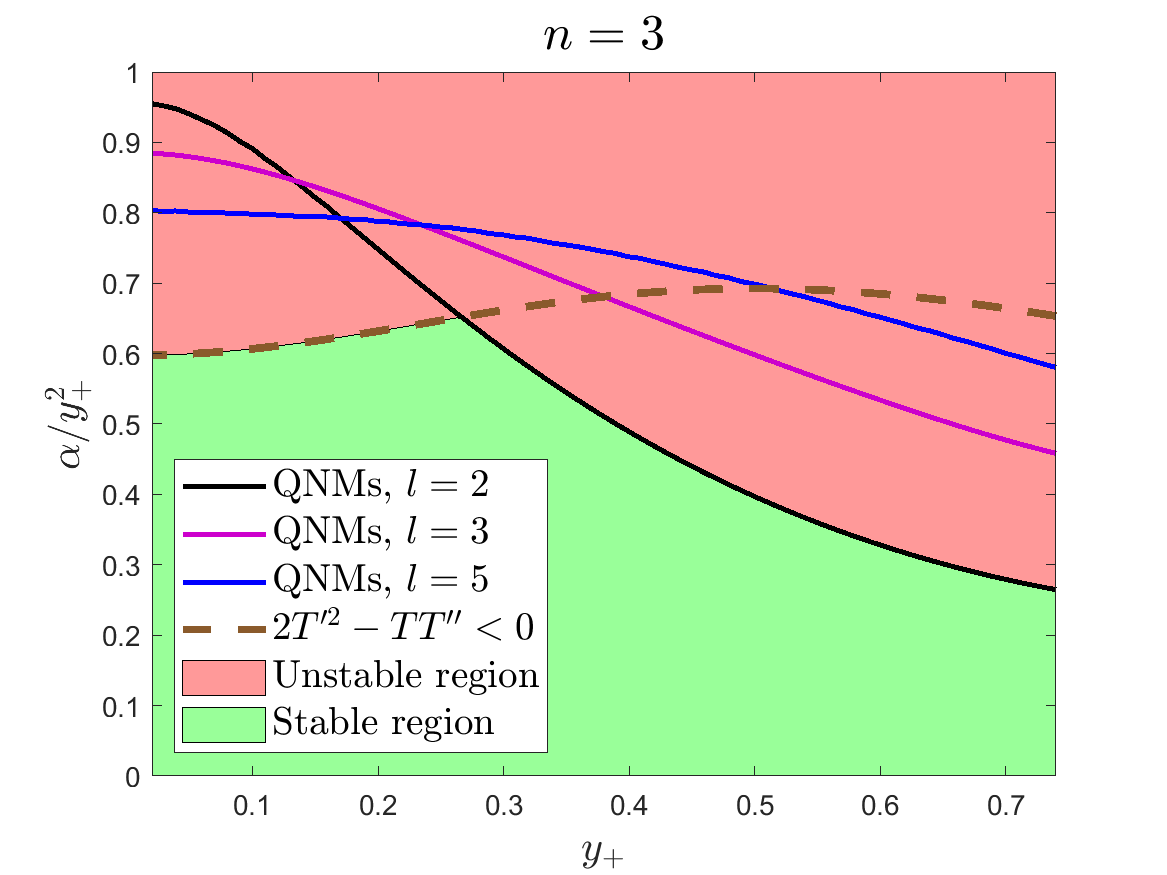}
			\caption{ \textbf{Left panel}: Stability and instability regions for higher multipole number $l$ for $n=3$ at extremality. For each $l$, the onset of instability is denoted by a solid curve above which the background becomes dynamically unstable. The dashed brown line is from the criterion~\eqref{Scriterion} for $l\rightarrow\infty$ most unstable mode.  \textbf{Right panel}: Stability and instability regions in terms of $y_+$ and $\alpha/y_+^2$.}\label{fig:highl_phase}
		\end{center}
	\end{figure*}

	In Fig.~\ref{fig:highl_phase}, we show the parameter space after considering the eikonal instability. For a given $l$, we use a solid curve to denote the critical line that separates the stability and instability regions. While the instability is dominated by the $l=2$ multipole for large $y_+$, the eikonal instability controls the instability for small $y_+$ region. Our numerics shows that the critical line behaves as $\alpha\sim y_+^2$ for small $y_+$. To see the contribution from higher $l$ modes clearly, we use $\alpha/y_+^2$ instead of $\alpha$ in the right panel of Fig.~\ref{fig:highl_phase}. It is manifest that the upper left corner is determined by the eikonal instability in the $y_+$-$\alpha/y_+^2$ plane. We find that the amount of computation increases rapidly with the multipole number. Therefore, we are not able to fix the critical line when $l>5$ due to the limitation of computational resources. Nevertheless, our numerical analysis suggests that the parameter space for stability and instability regions will not change qualitatively even after considering much higher $l$.
	
	To fix the instability region for the $l\rightarrow\infty$ most unstable mode, in particular, at small $y_+$, we incorporate the criteria for stability by the S-deformation approach~\cite{Ishibashi:2003ap,Kodama:2003ck}. The main idea is given as follows. For a Shr\"{o}dinger type equation $\mathcal{H}\Psi=\omega^2\Psi$, we define the inner product as
	\begin{equation}
		(\Psi, \Psi)=\int \Psi^\dagger(x)\Psi(x)dx\,,
	\end{equation}
	with $``^\dagger"$ the complex conjugation. Then there is the inequality
	\begin{equation}
		(\Psi_0,\mathcal{H} \Psi_0)\geqslant \omega_0^2 (\Psi_0, \Psi_0)\,,
	\end{equation}
	where $\Psi_0$ is an arbitrary smooth function with compact supports and $\omega_0^2$ is the lower bound of spectra. This inequality suggests that there exist instabilities if one can find a trial function $\Psi_0$ such that $(\Psi_0,\mathcal{H} \Psi_0)$ is negative. 

	For the master equations~\eqref{Meq}, it was found in~\cite{Takahashi:2012np} that if there is a region in which 

	\begin{equation}\label{Scriterion}
		\frac{dT}{dr}>0,\quad   2\left(\frac{dT}{dr}\right)^2 -T \frac{d^2T}{dr^2}<0\,,
	\end{equation}
	with the function $T$ defined in~\eqref{myT}, then there will be instability as $l\rightarrow\infty$. The resulted critical line from~\eqref{Scriterion} is shown by the dashed line in Fig.~\ref{fig:highl_phase}. As expected, we find that the instability region becomes large at small $y_+$. The overlap of regions of all types of instabilities produces the instability region for $n=3$ case (the green region of Fig.~\ref{fig:highl_phase}). For other spacetime dimensions, we check that there is no region in which the instability criterion~\eqref{Scriterion} can be satisfied. So we are not able to say anything for the instability with sufficiently large $l$ modes from~\eqref{Scriterion} in other dimensions. This is consistent with our observation that the instability is absent or suppressed for high $l$ modes for $n\geqslant 4$. Moreover, it is clear that the criterion~\eqref{Scriterion} is a sufficient but not necessary condition for the
	presence of dynamical instability.

	\subsection{Non-extremal black hole}\label{sec:Tneq0}
	Up until now, we have restricted ourselves to the extremal case with $Q/Q_{ext}=1$ and have confirmed that the $AdS_2$ BF bound violation observed in Fig.~\ref{fig:BF} provides a sufficient but not necessary criterion for the presence of an instability. By continuity such instability should extend away from extremity.  To confirm this point and to have a broader perspective of the instability, in this part we search directly for the unstable modes in the non-extremal GB-RNdS black hole. The non-extremal black hole is parametrized by three parameters $\{y_+,\alpha, q\}$. To obtain the full instability parameter space is beyond the scope of our present work. Let's consider two examples for $n=6$ and leave other cases for future study. 
	\begin{figure*}[ht]
		\centering
		\includegraphics[width=2.9in]{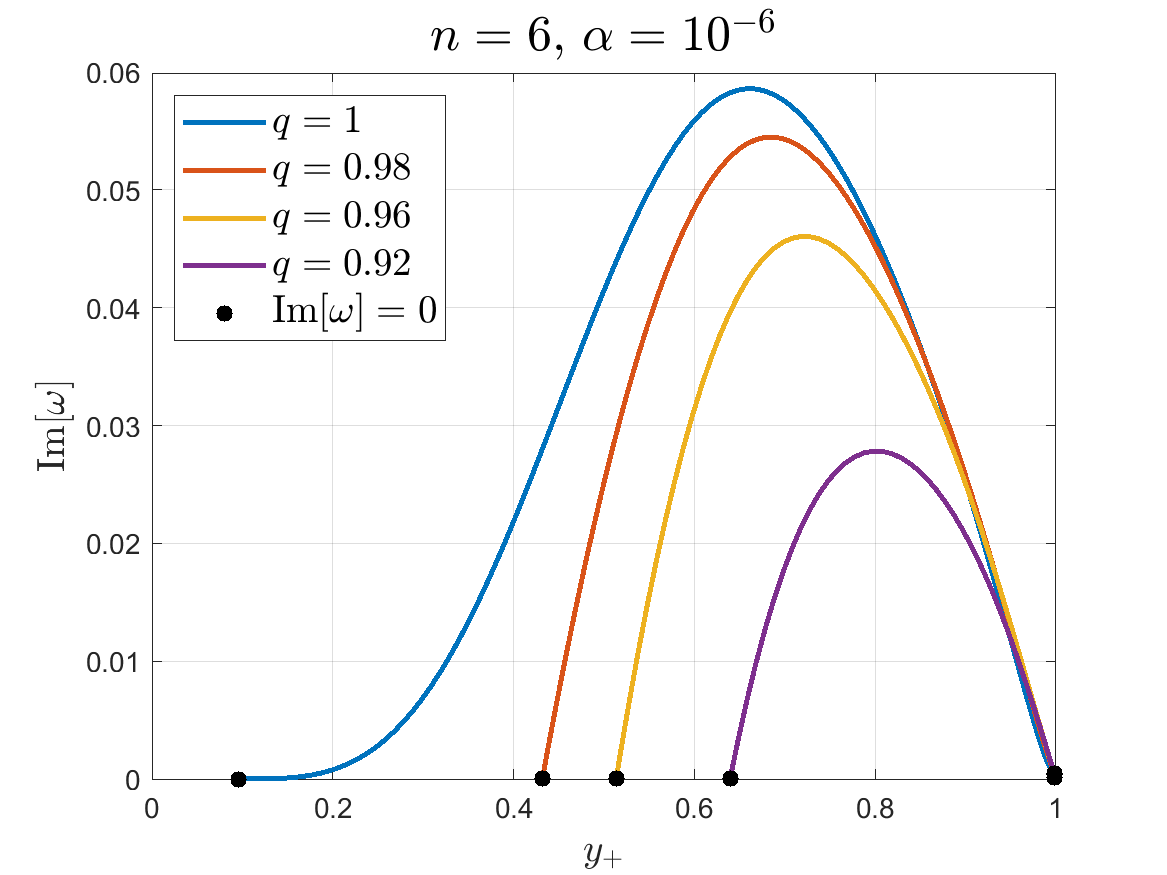}
		\includegraphics[width=2.9in]{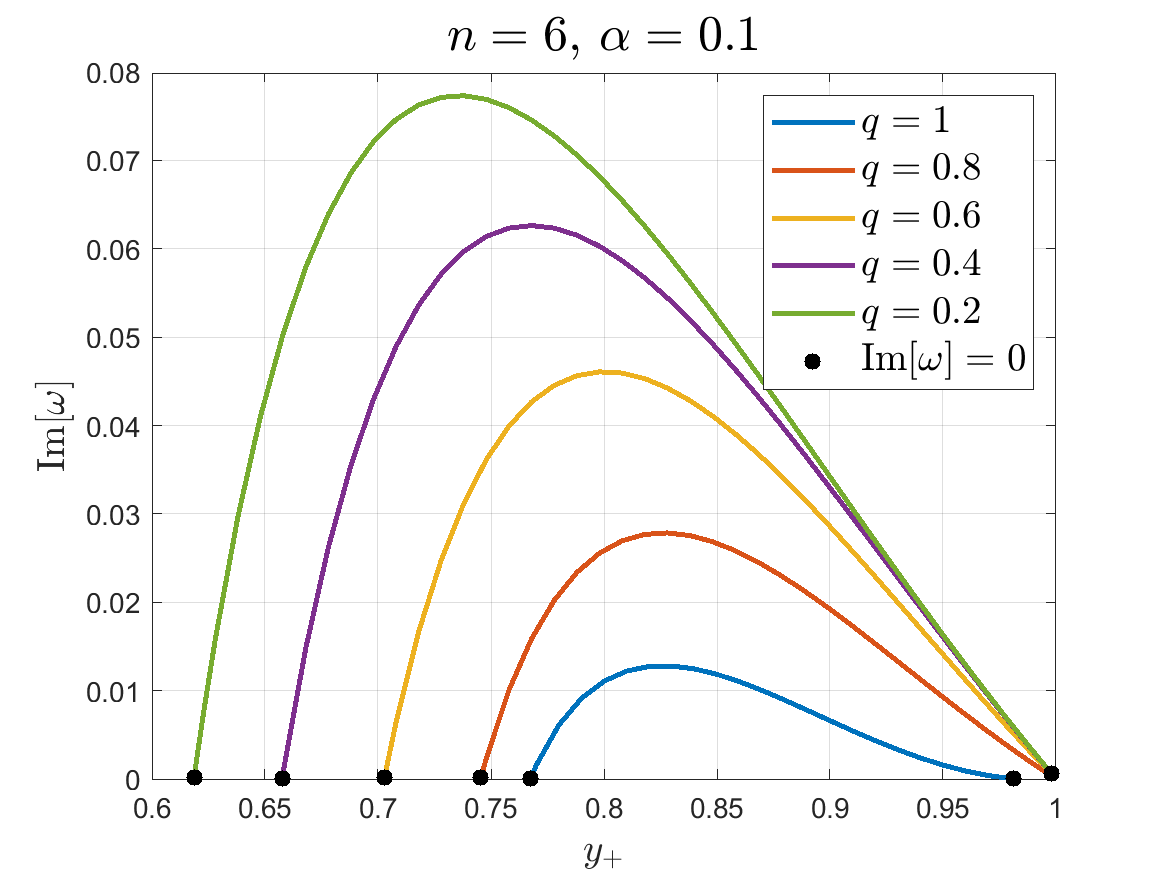}
		\caption{Instability timescale for non-extremal GB-RNdS black holes in eight dimensions ($n=6$) for different charge ratio $q=Q/Q_{ext}$. \textbf{Left panel}: The case for $\alpha=10^{-6}$. The extremal black hole are more unstable, which is similar the RNdS black hoe~\cite{Dias:2020ncd}. \textbf{Right panel}: The case with $\alpha=0.1$, for which the non-extremal black holes are found to be more unstable. The onset of instability at which \text{Im}$[\omega]=0$ is marked by black dots.}\label{region_n6_noex}
	\end{figure*}

	We present the instability timescale for $\alpha=10^{-6}$ and $\alpha=0.1$ for non-extremal GB-RNdS black holes in Fig.~\ref{region_n6_noex}. For a given charge away from extremality, the background is unstable for $y_+$ above a non-vanishing critical value. When $\alpha$ is small, the system becomes less unstable as the charge is moved away from extremality (see the left panel of Fig.~\ref{region_n6_noex}). This is similar to the RNdS case~\cite{Dias:2020ncd} corresponding to $\alpha\rightarrow 0$. On the contrary, for a larger GB coupling $\alpha=0.1$ in the right panel, the system becomes more unstable away from extremality. The above results uncover that the effect of GB coupling on the instability is pretty complicated, and also suggest that there might be new origin of the instability at finite temperature. 
	
	\section{Proof of non-oscillation of unstable modes}\label{sec:proof}
	Before concluding, we should mention that for the QNM frequencies with positive imaginary part, \emph{i.e.} Im$[\omega]>0$, we find numerically that their real part are vanishing (Re$[\omega]$=0). This is in agreement with the observation that unstable modes cannot be oscillatory when perturbing spherically symmetric static black holes~\cite{Konoplya:2008yy}. 
	
	The observation of only purely imaginary unstable modes can be understood as follows. We multiply the master equations~\eqref{Meq} by the complex conjugated functions $\phi_{\omega l}^\dagger$ and $A_{\omega l}^\dagger$, and then consider the integral of the resulted equation from the event horizon ($r_*\rightarrow-\infty$) to the cosmological horizon ($r_*\rightarrow+\infty$).
	\begin{align}
		I=&\int_{-\infty}^{\infty}\Big( \left[\phi_{\omega l}^\dagger, A_{\omega l}^\dagger\right]\frac{d^2}{dr_*^2}\begin{bmatrix}
			\phi_{\omega l}\\
			A_{\omega l}
		\end{bmatrix}+\omega^2\left[\phi_{\omega l}^\dagger, A_{\omega l}^\dagger\right]\begin{bmatrix}
			\phi_{\omega l}\nonumber\\
			A_{\omega l}
		\end{bmatrix}\\
	&-\left[\phi_{\omega l}^\dagger, A_{\omega l}^\dagger\right]\begin{bmatrix}
			V_g& V_c\\
			V_c& V_{em}
		\end{bmatrix}\begin{bmatrix}
			\phi_{\omega l}\\
			A_{\omega l}
		\end{bmatrix}\Big) dr_*\,.
	\end{align}
	Integration of the first term by parts yields
	\begin{align}\label{eqI}
			I=&\phi_{\omega l}^\dagger\frac{d\phi_{\omega l}}{dr_*}\Big{|}_{-\infty}^{\infty}+A_{\omega l}^\dagger\frac{dA_{\omega l}}{dr_*}\Big{|}_{-\infty}^{\infty}\nonumber\\
			&+\int_{-\infty}^{\infty}\Big{[} \omega^2(|\phi_{\omega l}|^2+|A_{\omega l}|^2)
			-\left(\Big{|}\frac{d\phi_{\omega l}}{dr_*}\Big{|}^2+\Big{|}\frac{dA_{\omega l}}{dr_*}\Big{|}^2\right)\nonumber\\
			&-\left(V_g|\phi_{\omega l}|^2+V_{em}|A_{\omega l}|^2+2V_c\text{Re}\left[\phi_{\omega l}^\dagger A_{\omega l}\right]\right)\Big{]}dr_*=0\,.
	\end{align}
	Note that we consider the ingoing boundary condition at the event horizon and the outgoing boundary condition at the cosmological horizon, (the boundary conditions are discussed in detail in Appendix~\ref{app:detail}, ) from which we have
	\begin{equation}\label{mybc}
		\begin{split}
			&\text{event horizon} \;r_*\rightarrow -\infty\,,\\
			&\frac{d\phi_{\omega l}}{dr_*}=-i\omega \phi_{\omega l}\,,\quad \frac{dA_{\omega l}}{dr_*}=-i\omega A_{\omega l}\,, \quad\\ 
			&\text{cosmological horizon} \;r_*\rightarrow \infty\,,\\
			&\frac{d\phi_{\omega l}}{dr_*}=i\omega \phi_{\omega l}\,,\quad\;\;\, \frac{dA_{\omega l}}{dr_*}=i\omega A_{\omega l}\,. \quad\\ 
		\end{split}\,
	\end{equation}
	Substituting above boundary conditions into~\eqref{eqI} and taking the imaginary part of the resulted equation, we obtain 
	\begin{equation}
		\begin{split}
			0&=\text{Im}[I]\\
			=&\text{Re}[\omega]\Big(|\phi_{\omega l}(\infty)|^2+|A_{\omega l}(\infty)|^2\\
			&+|\phi_{\omega l}(-\infty)|^2+|A_{\omega l}(-\infty)|^2\Big)\\
			&+2\text{Re}[\omega]\text{Im}[\omega]\int_{-\infty}^{\infty}(|\phi_{\omega l}|^2+|A_{\omega l}|^2) dr_*\,.
		\end{split}
	\end{equation}
	It is clear that Im$[\omega]<0$ for non-vanishing Re$[\omega]$. Therefore, the unstable modes (Im$[\omega]>0$) should not be oscillating, \emph{i.e.} Re$[\omega]=0$. 
	
	As a consequence, at the onset of instability there is a non-trivial static zero mode perturbation, suggesting the existence of a new non-spherically symmetric branch of solutions. We emphasize that the boundary conditions~\eqref{mybc} plays an important role in above proof. While it is natural to choose ingoing boundary condition at the event horizon, it is possible to consider other kinds of boundary conditions at the cosmological horizon, \emph{e.g.} the boundary condition with reflection and transmission for  black hole superradiance~\cite{Brito:2015oca}. For the later, one could have oscillating modes. On the other hand, the $AdS_2$ BF bound violation is local and is independent of any boundary condition.

	\section{Conclusion and discussion}\label{sec:discussion}
	In this work we have studied the instability of the GB-RNdS black hole~\eqref{RN} under gravito-electromagnetic perturbations. The scalar type perturbations are governed by two coupled second order differential equations~\eqref{Meq}. We have adopted two criteria to search for an instability. We have firstly performed a detailed analysis of the near-horizon limit of extremal GB-RNdS black holes and found that the scalar type modes behave as a massive scalar field in an $AdS_2$ background. According to the Durkee-Reall instability criterion~\cite{Durkee:2010ea,Hollands:2014lra}, the full extremal geometry will be unstable if the near-horizon effective mass violates the $AdS_2$ BF bound. Then, we have used full numerical analysis to obtain the behaviors of QNMs, from which the onset of the instability is identified by looking for a QNM frequency with a positive imaginary part. We found numerically that the real part of all unstable modes is vanishing, \emph{i.e.} Re$[\omega]=0$. Moreover, a religious proof was provided to show that the unstable modes should be purely imaginary.
	
	The instability of the black hole~\eqref{RN} has been found to depend not only on its size $y_+$ and charge ratio $q$, but also significantly on the GB coupling and the spacetime dimension. We have uncovered the gravitational instability in the GB-RNdS black hole when the spacetime dimension $d\geqslant 5$ ($n=d-2\geqslant 3$). The complete parameter space at extremality in terms of $y_+$ and $\alpha$ has been constructed, see Figs.~\ref{fig:region_n3},~\ref{fig:region_n56},~\ref{fig:region_n8} and~\ref{fig:highl_phase}. We have also provided strong numerical evidence for the Durkee-Reall instability criterion with positive cosmological constant $\Lambda>0$. While it is a sufficient criterion for the presence of instability at extremality, it is not a necessary one. By continuity, the instability extends away from extremelity. We have briefly discussed the unstable modes for the non-extremal GB-RNdS black hole. Depending on the value of GB coupling, the non-extremal solutions can be more stable or unstable than the extremal case.
	
	Compared with the RNdS case in the Einstein-Maxwell theory with a positive cosmological constant, some novel features have been found for the GB-RNdS black hole in our present work. 
	\begin{itemize}
		\item While it has been found that the RNdS black holes in $d=5$ are linearly-mode stable~\cite{Kodama:2003kk}, we have found that the presence of GB term makes the black holes to be unstable to gravitational perturbations in $d=5$ and higher dimensions. 
		
		\item For the five dimensional case, the instability occurs at $l=2$, but it becomes stronger at high multipole numbers $l$. Such eikonal instability was found to be closely related to the violation of the $AdS_2$ BF bound near the horizon of the extremal GB-RNdS black hole.
		
		\item It has been recently shown in~\cite{Dias:2020ncd} that at extremality the gravitational instability in the RNdS black hole is present when $d\geqslant 6$ for the whole range of $y_+=r_+/r_c$, \emph{i.e.} $0<y_+<1$, although typically the near-horizon $AdS_2$ BF bound is only violated for a finite range of $y_+$. It ceases to be valid for the charged black with GB term. Except for $d=5$ (see Figs.~\ref{fig:region_n3} and~\ref{fig:highl_phase}), the instability does not present in the full range $0<y_+<1$ (see \emph{e.g.} Figs.~\ref{fig:region_n56} and~\ref{fig:region_n8}).
		
		\item The physical origin of the instability of RNdS was argued to be due to the $AdS_2$ BF bound violation in the near-horizon limit of extremal background. It seems not to work for our GB case. The reasons are two-fold. Firstly, the instability in general is not present in the whole range $0<y_+<1$. Secondly, as can be seen \emph{e.g.} from Fig.~\ref{fig:region_n56}, for some choice of $\alpha$, there exists unstable region in which the $AdS_2$ BF bound is no longer violated for any $y_+$.
	\end{itemize}
{By using the large $d$ expansion, the GB-RNdS black hole was found to be unstable in the $d\rightarrow\infty$ limit, when the cosmological constant is sufficiently large~\cite{Chen:2017hwm}. Our findings are qualitatively consistent with the large $d$ results for higher-dimensional cases, $n\geqslant 7$. In the large $d$ discussion~\cite{Chen:2017hwm}, the local instability due to the BF bound violation near the extremal horizon regime was not considered. Moreover, compared to the large $d$ expansion, there are some novel features that only happen in lower dimensions, $n=5, 6$, see Figs.~\ref{fig:region_n56} and~\ref{fig:highl_phase}.}
	
	While the instability has been uncovered at the linear order, it is still an open question what is the nature of the endpoint of the instability.  If the onset of such instability leads to a new solution, it seems to be some yet undiscovered black holes with no spatial isometries as the modes triggering the instability are harmonics with $l\geqslant 2$. It will be quite interesting to construct these novel black holes numerically. Our numerical analysis suggests that the physical origin of the instability of the GB-RNdS black hole is not solely due to the violation of the $AdS_2$ BF bound. Further effort is needed to understand the physical origin of this instability. 
	
	The present work can be extended in several directions. We have mainly limited ourselves to the extremal case. It is interesting to study the QNM structure away from extremality, which is expected to exhibit intricate behaviors with bifurcations and merges as found in RNdS black holes~\cite{Dias:2020ncd}. We have focused on the instabilities triggered by the scalar type perturbations. To further understand the instability of the GB-RNdS black hole, one needs to consider the vector and tensor modes, as the black hole instability region is a combination of the instability regions of all types of perturbations. Another interesting extension is to consider the case with non-linear electrodynamics, in particular the DBI action that naturally arises in the low energy limit of string theory. Moreover, given the fact that near-extremal charge corresponds to fast rotating black holes, \emph{i.e.} the charge-angular momentum analogy, it will be interesting to extend the study to the rotating case. We leave these questions to future work.

	\section*{Acknowledgement}
	This work was supported in part by the National Key Research and Development Program of China Grant No.2020YFC2201501, by the National Natural Science Foundation of China Grants No.12122513, No.12075298, No.11991052 and No.12047503, and by the Key Research Program of the Chinese Academy of Sciences (CAS) Grant NO. XDPB15 and the CAS Project for Young Scientists in Basic Research YSBR-006.

	\appendix
	
	\section{Numerical details}\label{app:detail}

	We provide numerical details in this section. We will use the Chebyshev collocation scheme to solve above equations, see~\cite{Dias:2015nua} for a review on this method. 
	The Newton-Raphson Algorithm is then adopted to numerically solve $\{\omega,\phi_{\omega l},A_{\omega l}\}$. For a given initial value of $\{\omega,\phi_{\omega l},A_{\omega l}\}$, we stop the iteration when the error is small enough and we get the eigenvalue $\omega$ and the eigenfunction $\{\phi_{\omega l},A_{\omega l}\}$. We call the initial value a seed. 
	
	\subsection{Numerical setup}
	To compute the QNMs, we impose the ingoing boundary condition at the event horizon and the outgoing boundary condition at the cosmological horizon. These boundary conditions can be found as follows.
	
	Near a horizon, the master equations~\eqref{Meq} become
	\begin{align}
		\frac{\dif^2 \phi_{\omega l}}{\dif r_*^2}+\omega^2\phi_{\omega l}=0,\quad \frac{\dif^2 A_{\omega l}}{\dif r_*^2}+\omega^2A_{\omega l}=0\,,
	\end{align}
	where $\dif r_*=\dif r/f$. The solutions of these equations are given by
	\begin{align}\label{BChorizon}
		\phi_{\omega l}\approx e^{\pm{i \omega r_*}},\quad A_{\omega l} \approx e^{\pm{i \omega r_*}}\,.
	\end{align}
	Outgoing boundary condition at the cosmological horizon $r_c$ requires choosing the lower sign in~\eqref{BChorizon}, \emph{i.e.}
	\begin{align}
		\phi_{\omega l}\approx (r_c-r)^{-\frac{i \omega}{2\pi T_c}},\quad A_{\omega l}\approx (r_c-r)^{-\frac{i \omega}{2\pi T_c}}\,,
	\end{align}
	with both the cosmological temperature $T_c$ and the frequency $\omega$ measured in units of $r_c$. Similarly, ingoing boundary condition at the event horizon $r_+$ requires choosing the upper sign in~\eqref{BChorizon}, yielding
	\begin{align}
		\phi_{\omega l}\approx (r-r_+)^{{i \omega}\beta_1}e^{\frac{i\omega\beta_2}{r-r_+}},\quad A_{\omega l}\approx (r-r_+)^{{i \omega}\beta_1}e^{\frac{i\omega\beta_2}{r-r_+}},
	\end{align}
	with
	\begin{align}
		\beta_1=\frac{2f^{'''}(r_+)}{f^{''}(r_+)^2},\quad \beta_2=\frac{2}{f^{''}(r_+)}\,.
	\end{align}
	Here we have considered the extremal case with the event horizon temperature $T_+=0$.
	
	For numerical convenience, we perform the following field redefinition. 
	\begin{align}
		&\phi_{\omega l}(r)=B\tilde{\phi}_{\omega l}(r),\quad A_{\omega l}(r)=B\tilde{A}_{\omega l}(r)\,,\\
		&B=(r_c-r)^{-\frac{i \omega}{2\pi T_c}}(r-r_+)^{{i \omega}\beta_1}e^{\frac{i\omega\beta_2}{r-r_+}}\,,
	\end{align}
	such that $\tilde{\phi}_{\omega l}(r)$ and $\tilde{A}_{\omega l}(r)$ are smooth functions of $r$ with a regular Taylor series expansion at both event  and cosmological horizons.
	For the non-extremal black hole case, the function $B$ can be chosen to be
	\begin{align}
		B=(r_c-r)^{-\frac{i \omega}{2\pi T_c}}(r-r_+)^{-\frac{i \omega}{2\pi T_+}}.
	\end{align}

	To proceed, we introduce a new compact radial coordinate given by
	\begin{align}
		y=\left(1-\left(\frac{r_c-r}{r_c-r_+}\right)^{1/2}\right)^{1/\xi}\,,
	\end{align}
	where $\xi$ is a parameter for numerical simplicity and it controls grid density in numerical calculation. We will choose $\xi=2,4,8,16,32,64\dots$ for different cases in our numerics.\, {For example, the QNMs were numerically computed with $\xi=2$ for the RNdS case in~\cite{Dias:2020ncd}, and to compute the QNMs in Fig.~\ref{fig:critical_n4}, we will choose $\xi=96$.}
	In the new coordinate, $y = 0$ now describes the black hole event horizon ($r = r_+$), and $y = 1$ the cosmological horizon ($r = r_c$). {$\phi_{\omega l}$ and $A_{\omega l}$ in this coordinate become
		\begin{align}
			{\phi}_{\omega l}(y)=e^{\frac{i \lambda_1 \omega}{y^\xi (2 -  y^\xi)}}y^{i \xi \lambda_2 \omega/2}(1 -  y^\xi)^{- \frac{i \omega}{2\pi T_c}}\tilde{\phi}_{\omega l}(y)\,,\label{phi_ansatz}\\
			{A}_{\omega l}(y)=e^{\frac{i \lambda_1 \omega}{y^\xi (2 -  y^\xi)}}y^{i\xi \lambda_2 \omega/2}(1 -  y^\xi)^{- \frac{i \omega}{2\pi T_c}}\tilde{A}_{\omega l}(y)\,,\label{A_ansatz}
		\end{align}
		with
		\begin{align}
			\lambda_1=\frac{\beta_1}{1-y_+},\quad \lambda_2=\frac{\beta_2}{1-y_+}\,.
		\end{align}
		Finally, we can obtain the master equations in the new coordinate systems in terms of $\tilde{\phi}_{\omega l}(y)$ and $\tilde{A}_{\omega l}(y)$, for which the boundary conditions at both horizons are Neumann type, \emph{i.e.} $\partial_y\tilde{\phi}_{\omega l}(y)|_{y=0,1}=0$ and $\partial_y\tilde{A}_{\omega l}(y)|_{y=0,1}=0$. A discrete set of complex numbers of $\omega$ allowing non-trivial solutions for above system are known as QMN frequencies. 

		\begin{figure*}
			\centering
			\begin{minipage}{0.48\textwidth}
				\centering
				\includegraphics[width=\textwidth]{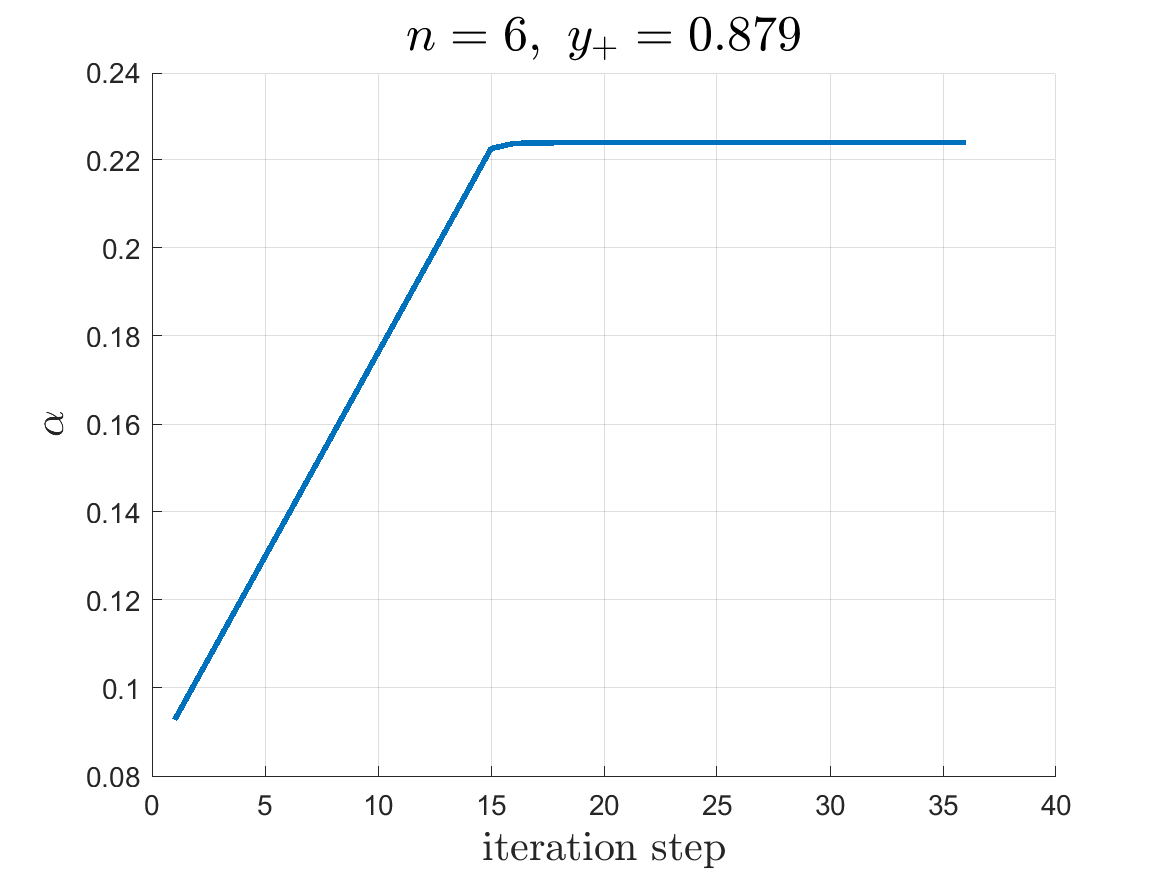}
			\end{minipage}
			\begin{minipage}{0.48\textwidth}
				\centering
				\includegraphics[width=\textwidth]{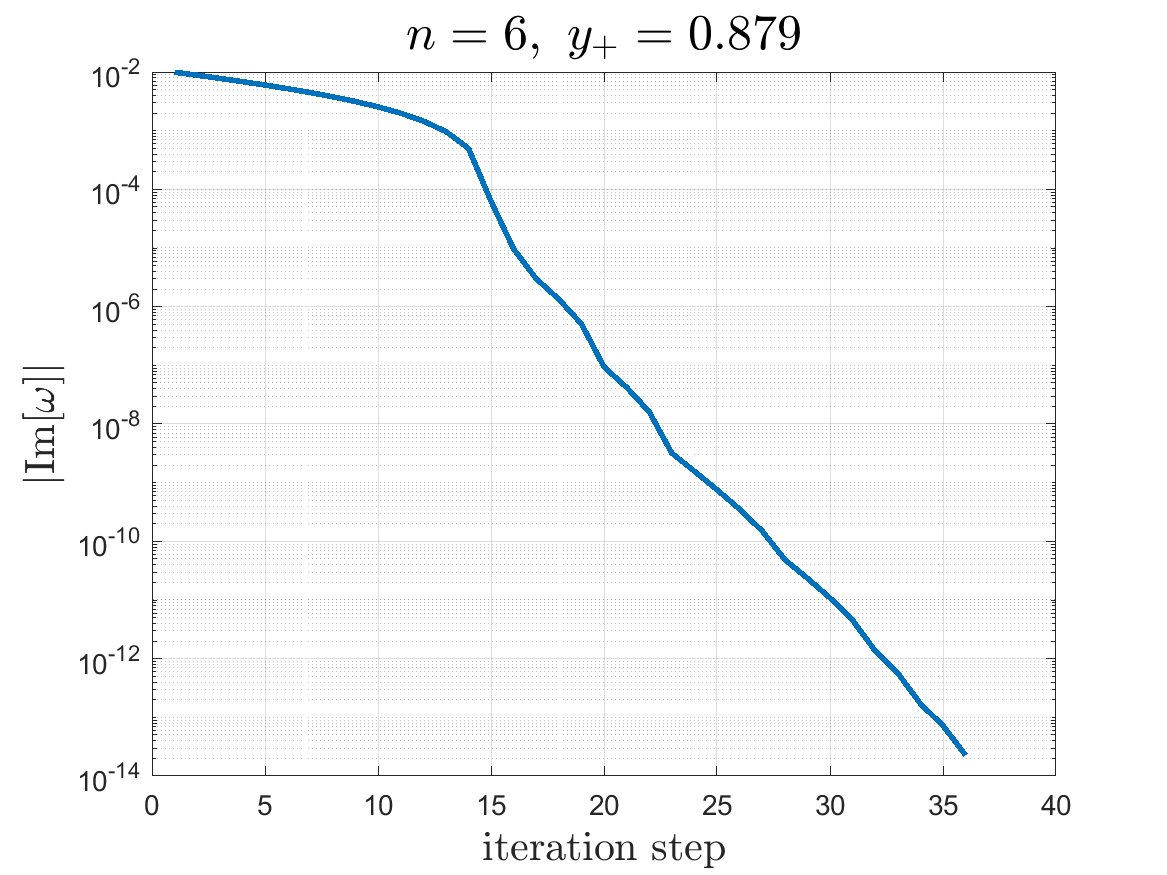}
			\end{minipage}
			
			\begin{minipage}{0.48\textwidth}
				\centering
				\includegraphics[width=\textwidth]{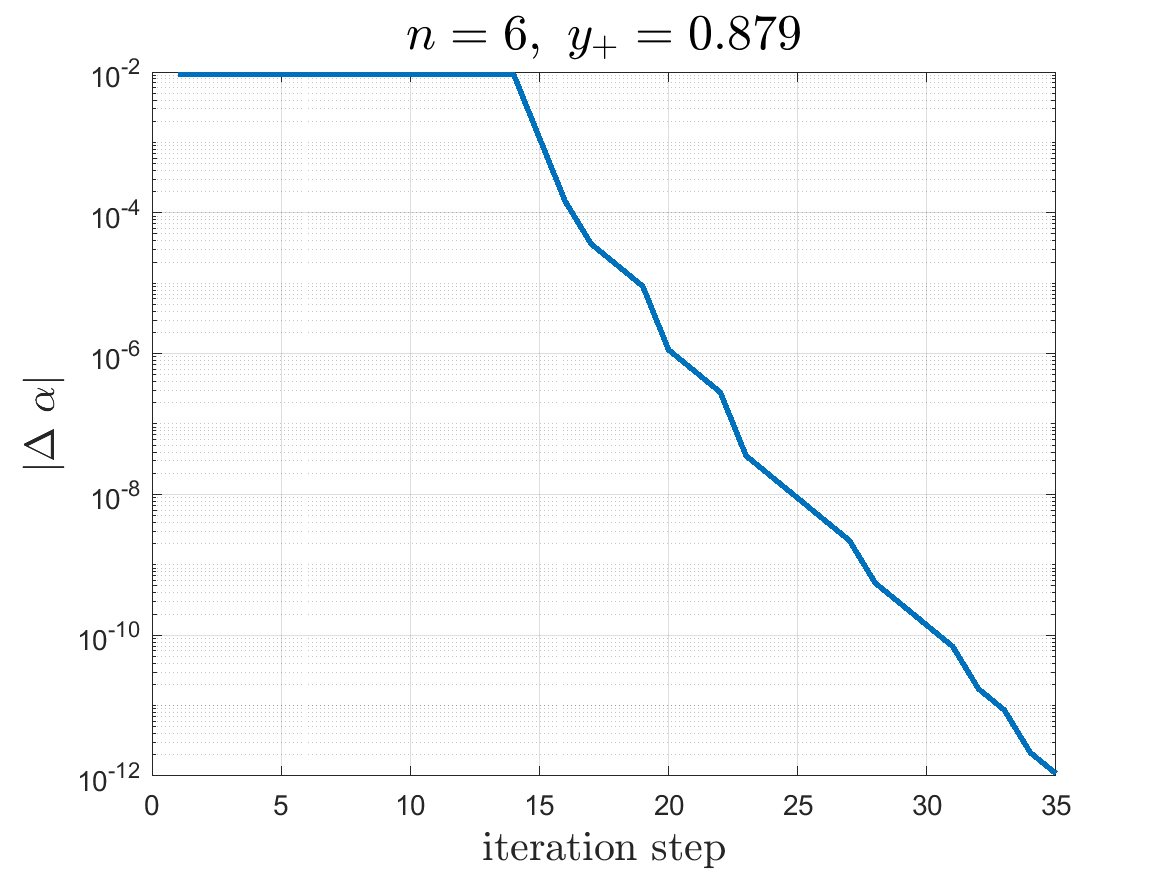}
			\end{minipage}
			\begin{minipage}{0.48\textwidth}
				\centering
				\includegraphics[width=\textwidth]{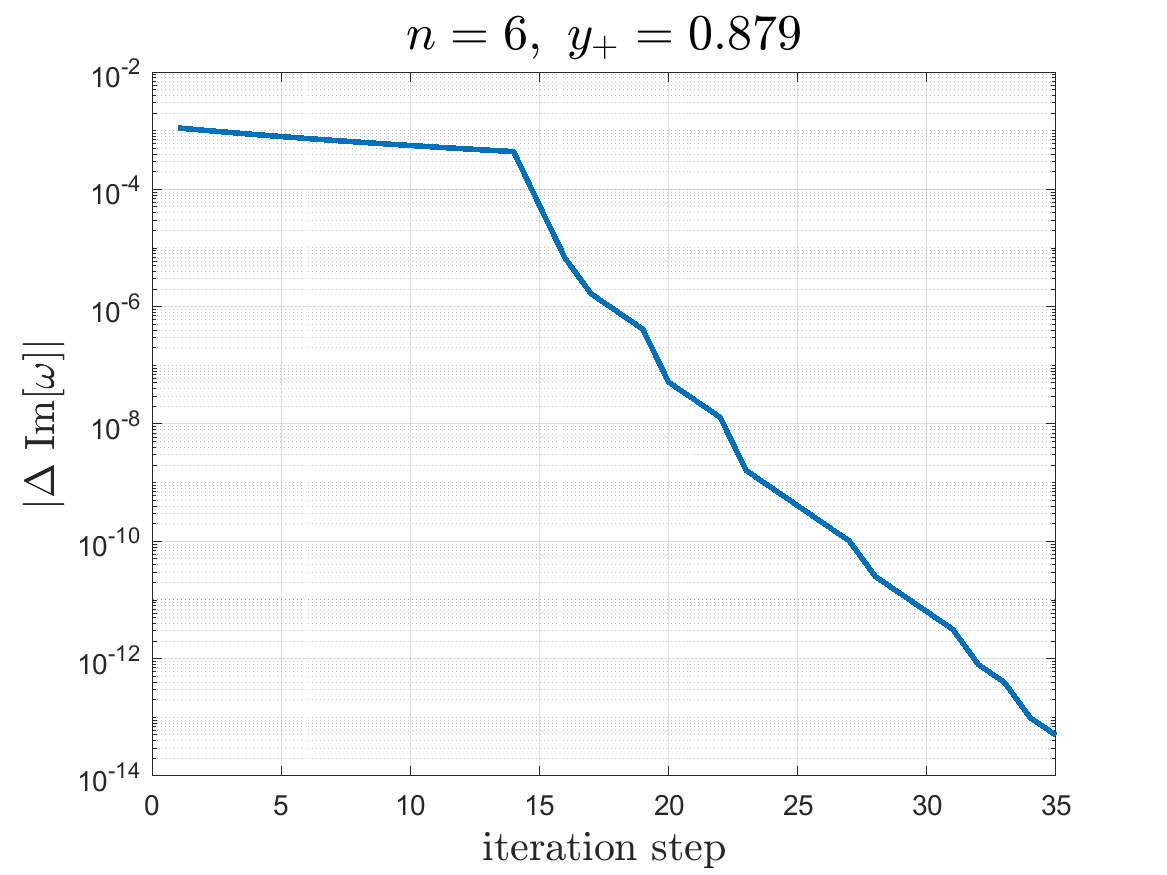}
			\end{minipage}
			\caption{Converging process for $y_+=0.879$ and $n=6$. We adjust $\alpha$ to get the critical point at which $\mathrm{Im}[\omega]=0$.}\label{n6_conv}
		\end{figure*}

		\subsection{Algorithm for the onset of instability}
		To obtain the stability and instability regions in the $y_+$-$\alpha$ plane for perturbations, it is crucial to find the onset of the instability at which \text{Im}$[\omega]=0$. Our algorithm for searching for the onset of instability is given as follows.

		\begin{figure*}
			\centering
			\includegraphics[width=0.6\textwidth]{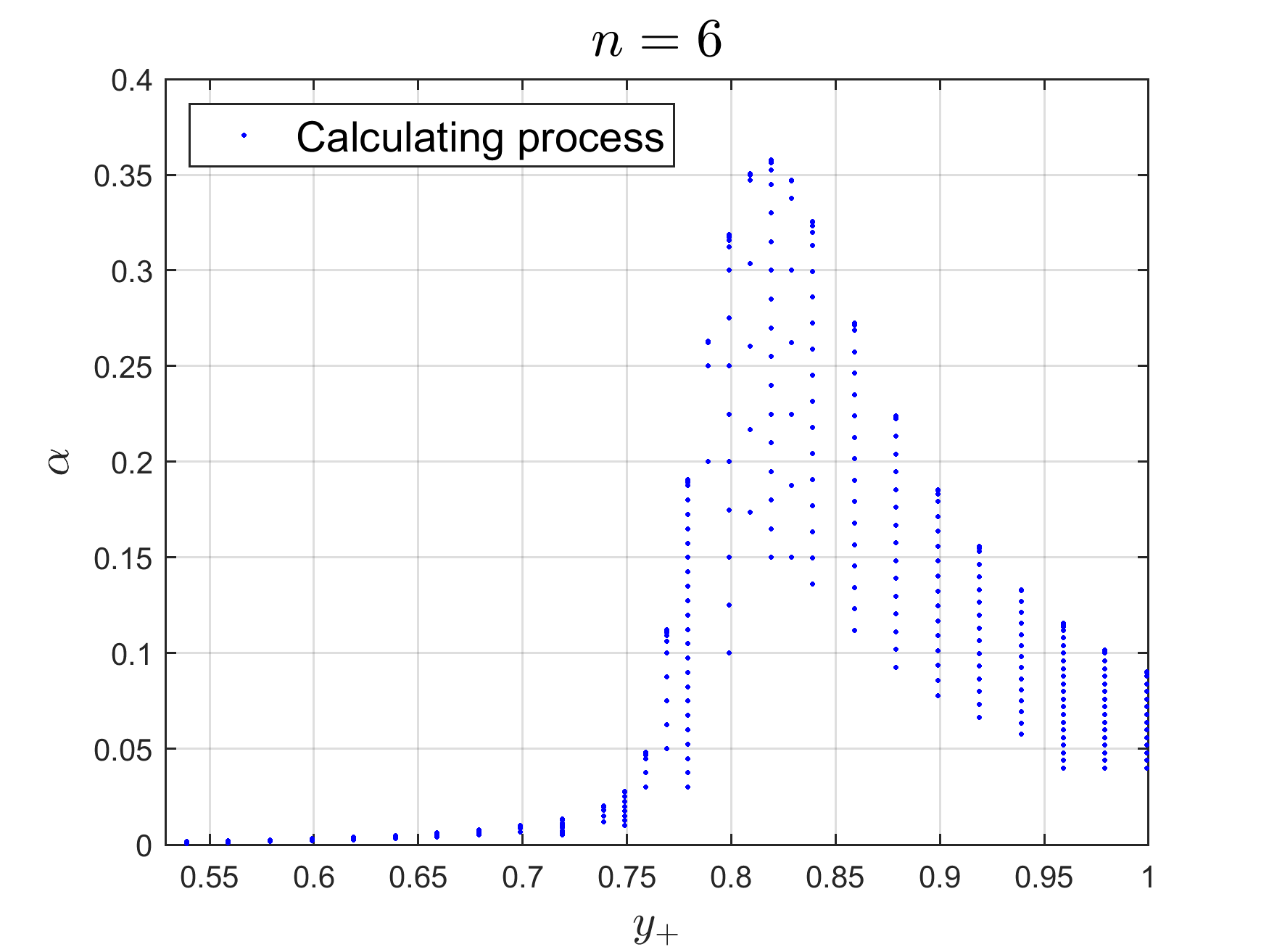}
			\caption{Parameter space that we have actually scanned for $n=6$.}\label{ND_n6_region}
		\end{figure*}

		In the $\{y_+,\alpha\}$ plane, we start from some point that is unstable (we can choose a pair of $\{y_+,\alpha\}$ for which the $AdS_2$ BF bound is violated). We adjust $\alpha$ to be $\alpha+\Delta\alpha$ with $y_+$ fixed, and use the former eigenvalue $\omega$ and eigenfunction as a seed to calculate the next eigenvalue and eigenfunction. If the iteration does not converge or converges to a $\omega$ with negative imaginary part, we will reduce the step on $\alpha$, $\Delta\alpha\rightarrow\Delta\alpha/2$, until $\Delta\alpha$ is smaller than some critical value. Our criterion for the break of  iteration is $\Delta\alpha<10^{-12}$. This process is illustrated in Fig.~\ref{n6_conv} for $y_+=0.879$ and $n=6$. One can also fix $\alpha$ first and adjust $y_+$ in a similar way, which is used to obtain the onset of instability (black dots) in Figs.~\ref{fig:alpha_n5},~\ref{fig:QNM_n8} and~\ref{region_n6_noex} in the main text.

		Then we choose a new $y_+\rightarrow y_++\Delta y_+$ with $\alpha\rightarrow\alpha+\Delta\tilde{\alpha}$. The adjustment can be fixed as follows. We compute the case with $y_++\Delta y_+$ by using the former eigenvalue $\omega$ and eigenfunction as a seed. If the iteration is converged and the imaginary part of $\omega$ is positive, we adjust $\alpha$ in the manner we introduce above. If the iteration does not converge or converges to a $\omega$ with negative imaginary part, we reduce the step on $y_+$, $\Delta y_+\rightarrow\Delta y_+/2$, until $\Delta y_+$ is smaller than a critical value, for example $\Delta y_+<10^{-10}$.  $\Delta\tilde{\alpha}$ is chosen by some test calculation. For example, if we start from $\{y_+=0.8,\alpha=0.32\}$ and try to calculate the case with $y_+<0.8$, $\Delta\tilde{\alpha}$ can be chosen to be $-0.2\alpha$. Because the computation of $\mathrm{Im}[\omega]$ is time consuming, in practice we only need to scan the region near the onset of instability. The actual region of $\{y_+,\alpha\}$ we calculated is shown in Fig.~\ref{ND_n6_region}.

		After discretization the master equation by using the Chebyshev collocation scheme, the computation of the eigenvalue needs to get an inverse of the resulted matrix. It turns out that the matrix is near singular, so one needs sufficiently high precision to get the eigenvalue. Typically, we found 100 digits to be sufficient for our computation. For some special cases, we need 650 digits. So the numerical computation is very time consuming.



\begin{thebibliography}{30}
			\setlength{\itemsep}{2mm}
			\bibitem{Kodama:2003kk}
			H.~Kodama and A.~Ishibashi,
			``Master equations for perturbations of generalized static black holes with charge in higher dimensions,''
			Prog. Theor. Phys. \textbf{111}, 29-73 (2004)
			[arXiv:hep-th/0308128 [hep-th]].
			
			\bibitem{Konoplya:2008au}
			R.~A.~Konoplya and A.~Zhidenko,
			``Instability of higher dimensional charged black holes in the de-Sitter world,''
			Phys. Rev. Lett. \textbf{103}, 161101 (2009)
			[arXiv:0809.2822 [hep-th]].
			
			\bibitem{Konoplya:2013sba}
			R.~A.~Konoplya and A.~Zhidenko,
			``Instability of D-dimensional extremally charged Reissner-Nordstrom(-de Sitter) black holes: Extrapolation to arbitrary D,''
			Phys. Rev. D \textbf{89}, no.2, 024011 (2014)
			[arXiv:1309.7667 [hep-th]].
			
			\bibitem{Dias:2020ncd}
			O.~J.~C.~Dias and J.~E.~Santos,
			``Origin of the Reissner-Nordstr\"om\textendash{}de Sitter instability,''
			Phys. Rev. D \textbf{102}, no.12, 124039 (2020)
			[arXiv:2005.03673 [hep-th]].
			
			\bibitem{Cai:2003gr}
			R.~G.~Cai and Q.~Guo,
			``Gauss-Bonnet black holes in dS spaces,''
			Phys. Rev. D \textbf{69}, 104025 (2004)
			[arXiv:hep-th/0311020 [hep-th]].
			
			\bibitem{Cuyubamba:2016cug}
			M.~A.~Cuyubamba, R.~A.~Konoplya and A.~Zhidenko,
			``Quasinormal modes and a new instability of Einstein-Gauss-Bonnet black holes in the de Sitter world,''
			Phys. Rev. D \textbf{93}, no.10, 104053 (2016)
			[arXiv:1604.03604 [gr-qc]].
			
			\bibitem{Chen:2017hwm}
			B.~Chen and P.~C.~Li,
			``Static Gauss-Bonnet Black Holes at Large $D$,''
			JHEP \textbf{05}, 025 (2017)
			[arXiv:1703.06381 [hep-th]].
			
			\bibitem{Durkee:2010ea}
			M.~Durkee and H.~S.~Reall,
			``Perturbations of near-horizon geometries and instabilities of Myers-Perry black holes,''
			Phys. Rev. D \textbf{83}, 104044 (2011)
			[arXiv:1012.4805 [hep-th]].
			
			
			
			
			\bibitem{Wiltshire:1985us}
			D.~L.~Wiltshire,
			``Spherically Symmetric Solutions of Einstein-maxwell Theory With a {Gauss-Bonnet} Term,''
			Phys. Lett. B \textbf{169}, 36-40 (1986)
			
			\bibitem{Cai:2001dz}
			R.~G.~Cai,
			``Gauss-Bonnet black holes in AdS spaces,''
			Phys. Rev. D \textbf{65}, 084014 (2002)
			[arXiv:hep-th/0109133 [hep-th]].
			
			
			\bibitem{Takahashi:2011qda}
			T.~Takahashi,
			``Fatal Effects of Charges on Stability of Black Holes in Lovelock Theory,''
			Prog. Theor. Phys. \textbf{125}, 1289-1310 (2011)
			[arXiv:1102.1785 [gr-qc]].
			
			\bibitem{Takahashi:2012np}
			T.~Takahashi,
			``Instability of Charged Lovelock Black Holes: Vector Perturbations and Scalar Perturbations,''
			PTEP \textbf{2013}, 013E02 (2013)
			[arXiv:1209.2867 [gr-qc]].
			
			\bibitem{Ishibashi:2003ap}
			A.~Ishibashi and H.~Kodama,
			``Stability of higher dimensional Schwarzschild black holes,''
			Prog. Theor. Phys. \textbf{110}, 901-919 (2003)
			[arXiv:hep-th/0305185 [hep-th]].
			
			\bibitem{Kodama:2003ck}
			H.~Kodama and A.~Ishibashi,
			``Stability of generalized static black holes in higher dimensions,''
			[arXiv:gr-qc/0312012 [gr-qc]].
			
			
			
			
			\bibitem{Hollands:2014lra}
			S.~Hollands and A.~Ishibashi,
			``Instabilities of extremal rotating black holes in higher dimensions,''
			Commun. Math. Phys. \textbf{339}, no.3, 949-1002 (2015)
			[arXiv:1408.0801 [hep-th]].
			
			\bibitem{Konoplya:2011qq}
			R.~A.~Konoplya and A.~Zhidenko,
			``Quasinormal modes of black holes: From astrophysics to string theory,''
			Rev. Mod. Phys. \textbf{83}, 793-836 (2011)
			[arXiv:1102.4014 [gr-qc]].
			
			\bibitem{Konoplya:2008yy}
			R.~A.~Konoplya, K.~Murata, J.~Soda and A.~Zhidenko,
			``Looking at the Gregory-Laflamme instability through quasi-normal modes,''
			Phys. Rev. D \textbf{78}, 084012 (2008)
			[arXiv:0807.1897 [hep-th]].
			
			\bibitem{Brito:2015oca}
			R.~Brito, V.~Cardoso and P.~Pani,
			``Superradiance: New Frontiers in Black Hole
			Physics,''
			Lect. Notes Phys. \textbf{906}, pp.1-237 (2015)
			[arXiv:1501.06570 [gr-qc]].
			
			
			\bibitem{Dias:2015nua}
			\'O.~J.~C.~Dias, J.~E.~Santos and B.~Way,
			``Numerical Methods for Finding Stationary Gravitational Solutions,''
			Class. Quant. Grav. \textbf{33}, no.13, 133001 (2016)
			[arXiv:1510.02804 [hep-th]].
			
		\end{thebibliography}
	\end{document}